# Roadmap on Atomic-scale Semiconductor Devices

Steven R. Schofield[1,2,28], Andrew J. Fisher[1], Eran Ginossar[3], Joseph W. Lyding[4], Richard Silver[5], Fan Fei[5], Pradeep Namboodiri[5], Jonathan Wyrick[5], M.G. Masteghin[6], D.C. Cox[6], B.N. Murdin[6], S.K Clowes[6], Joris G. Keizer[7,9,10,28], Michelle Y. Simmons[7,9,10,28], Holly G. Stemp[8,10], Andrea Morello[8,10], Benoit Voisin[7,9,10], Sven Rogge[7,10], Robert A. Wolkow[11], Lucian Livadaru[11], Jason Pitters[12], Taylor J. Z. Stock[1,13], Neil J. Curson[1,13,28], Robert. E. Butera[14], Tatiana V. Pavlova[15], A.M. Jakob[10,16], D. Spemann[17], P. Räcke[17,18], F. Schmidt-Kaler[19], D.N. Jamieson[10,16], Utkarsh Pratiush[20], Gerd Duscher[20], Sergei V. Kalinin[20], Dimitrios Kazazis[21], Procopios Constantinou[21], Gabriel Aeppli[21-24], Yasin Ekinci[21], James H.G. Owen[25,28], Emma Fowler[26], S. O. Reza Moheimani[26], John N. Randall[25,28], Shashank Misra[27], Jeffrey Ivie[27], Christopher R. Allemang[27], Evan M. Anderson[27], Ezra Bussmann[27], Quinn Campbell[27], Xujiao Gao[27], Tzu-Ming Lu[27] and Scott W. Schmucker[27]

[1]London Centre for Nanotechnology, University College London, 17-19 Gordon St, WC1H 0AH, UK
[2]Department of Physics and Astronomy, University College London, WC1E 6BT, UK
[3]School of Mathematics and Physics, University of Surrey, UK
[4]Department of Electrical and Computer Engineering, University of Illinois, Urbana, Illinois 61801, USA
[5]Atom Scale Device Group, National Institute of Standards and Technology, Gaithersburg, MD 20899, USA
[6]Advanced Technology Institute, University of Surrey, GU2 7XH, UK
[7]School of Physics, University of New South Wales Sydney, Australia
[8]School of Electrical Engineering & Telecommunications, UNSW Sydney, Australia
[9]Silicon Quantum Computing Pty Ltd, Sydney, Australia
[10]ARC Centre of Excellence for Quantum Computation and Communication Technology (CQC2T), Australia
[11]University of Alberta, Department of Physics, Edmonton, Canada T6G 2E1
[12]National Research Council of Canada, Quantum and Nanotechnologies Research Centre, 11421 Saskatchewan Drive NW. Edmonton Canada T6G 2M9.
[13]Department of Electronic and Electrical Engineering, University College London, WC1E 7JE, UK
[14]Laboratory for Physical Sciences, College Park, University of Maryland, MD, 20740, United States
[15]Prokhorov General Physics Institute of the Russian Academy of Sciences, Vavilov str. 38, 119991 Moscow, Russia
[16]School of Physics, The University of Melbourne, 3010 Parkville, VIC, Australia
[17]Leibniz-Institute of Surface Engineering (IOM), 04318 Leipzig, Germany
[18]Universität Leipzig, Felix Bloch Institute for Solid State Physics, Applied Quantum Systems, 04103 Leipzig, Germany
[19]QUANTUM, Institut für Physik, Johannes Gutenberg-Universität Mainz, 55128 Mainz, Germany
[20]Department of Materials Science and Engineering, University of Tennessee, Knoxville, TN 37996, USA
[21]Paul Scherrer Institute, 5232 Villigen, Switzerland
[22]Institute of Physics, Ecole Polytechnique Fédérale de Lausanne (EPFL), 1015 Lausanne, Switzerland
[23]Department of Physics, ETH Zürich, 8093 Zürich, Switzerland
[24]Quantum Center, Eidgenössische Technische Hochschule Zürich (ETHZ), 8093 Zürich, Switzerland
[25]Zyvex Labs, 1301 N. Plano Road, Richardson Texas, 75081 USA

[26]University of Texas at Dallas, Texas, United States of America
[27]Sandia National Laboratories, 1515 Eubank Blvd. SE, Albuquerque, New Mexico 87185, United States of America
[28]Guest Editor of the Roadmap

Guest Editor emails: s.schofield@ucl.ac.uk; n.curson@ucl.ac.uk; jowen@zyvexlabs.com; jrandall@zyvexlabs.com; j.g.keizer@unsw.edu.au; michelle.simmons@unsw.edu.au

**Contents:**

1. Introduction to semiconducting atomic devices – Steven R. Schofield, Neil J. Curson, James H. G. Owen, John N. Randall, Joris G. Keizer and Michelle Y. Simmons

2. Fundamentals
    2.1 Shallow dopants in silicon; theory – Andrew. J Fisher and Eran Ginossar
    2.2 Hydrogen Depassivation Lithography – Joesph Lyding
    2.3 P in Si device fabrication – Richard Silver, Fan Fei, Pradeep Namboodiri and Jonathan Wyrick
    2.4 Focussed ion beam as a tool to realize scalable one-million qubit devices – M.G. Masteghin, D.C. Cox, B.N. Murdin and S.K Clowes

3. Devices and applications
    3.1 Atomic-Precision Engineering of Atom Qubits for Quantum Computing – Joris G. Keizer and Michelle Y. Simmons
    3.2 Quantum control of donor spins – Holly G. Stemp and Andrea Morello
    3.3 In-situ quantum measurements – Benoit Voisin and Sven Rogge
    3.4 Silicon Dangling Bond Quantum Dot Devices – Robert A. Wolkow, Lucian Livadaru and Jason Pitters

4. Deterministic doping and alternative materials
    4.1 Deterministic dopant incorporation – Taylor J. Z. Stock, Steven R. Schofield and Neil J. Curson
    4.2 Tip-induced deterministic P incorporation – Jonathan Wyrick, Pradeep Namboodiri, Fan Fei and Richard Silver
    4.3 Alternative Precursors for Semiconducting Atomic-scale Devices – Robert. E. Butera
    4.4 Halogen resists – Tatiana V. Pavlova

5. Complementary approaches and scale-up
    5.1 Novel methods in ion implantation – A.M. Jakob, D. Spemann, P. Räcke, F. Schmidt-Kaler and D.N. Jamieson
    5.2 Atomic Fabrication by Electron Beams – Utkarsh Pratiush, Gerd Duscher and Sergei V. Kalinin
    5.3 Quantum device fabrication with EUV lithography – Dimitrios Kazazis, Procopios Constantinou, Gabriel Aeppli and Yasin Ekinci

**Abstract:**

Spin states in semiconductors provide exceptionally stable and noise-resistant environments for qubits, positioning them as optimal candidates for reliable quantum computing technologies. The proposal to use nuclear and electronic spins of donor atoms in silicon, introduced by Kane in 1998, sparked a new research field focused on the precise positioning of individual impurity atoms for quantum devices, utilising scanning tunnelling microscopy and ion implantation. This roadmap article reviews the advancements in the 25 years since Kane's proposal, the current challenges, and the future directions in atomic-scale semiconductor device fabrication and measurement. It covers the quest to create a silicon-based quantum computer and expands to include diverse material systems and fabrication techniques, highlighting the potential for a broad range of semiconductor quantum technological applications. Key developments include phosphorus in silicon devices such as single-atom transistors, arrayed few-donor devices, one- and two-qubit gates, three-dimensional architectures, and the development of a toolbox for future quantum integrated circuits. The roadmap also explores new impurity species like arsenic and antimony for enhanced scalability and higher-dimensional spin systems, new chemistry for dopant precursors and lithographic resists, and the potential for germanium-based devices. Emerging methods, such as photon-based lithography and electron beam manipulation, are discussed for their disruptive potential. This roadmap charts the path toward scalable quantum computing and advanced semiconductor quantum technologies, emphasising the critical intersections of experiment, technological development, and theory.

# Section 1 – Introduction to the Roadmap on Atomic-scale Semiconductor Devices

Steven R. Schofield, Neil J. Curson, James H. G. Owen, John N. Randall, Joris G. Keizer, and Michelle Y. Simmons

We can now construct electronic devices with active components made of precisely placed single atoms. This transformation has been driven by advancements in single-atom positioning using scanning tunnelling microscopy (STM) and ion implantation, alongside improvements in analytical tools, device measurement techniques, and theoretical methods. The grand challenge is to develop large-scale coupled arrays of impurity atoms in semiconductor hosts to create novel impurity-based devices for quantum information technology. Furthermore, the deterministic positioning of individual impurity atoms in semiconductors is opening new avenues for exploring the fundamental physics of matter.

Spin states in semiconductors provide one of the most stable and noise-resistant environments for qubits yet discovered, making them ideal hosts for developing reliable quantum computing technologies. Silicon offers the potential to integrate quantum and conventional semiconductor technologies within a single material system, promising a scalable quantum technology—a goal that has so far eluded all quantum computer architectures. The groundbreaking proposal to use nuclear and electronic spins of donor atoms in silicon to build a quantum computer was made by Kane in 1998. Although met with enthusiasm, his proposal faced a significant challenge: the need to create a large array of regularly spaced phosphorus impurity atoms in a silicon matrix, for which no known technology existed at the time.

This roadmap article presents an overview of the remarkable advancements made in atomic-scale semiconductor devices in the quarter-century since Kane's proposal. It will detail the current challenges and the scientific and technological developments necessary to overcome them. We will explore not just the ongoing quest to create a silicon-based quantum computer, but also the expansion of this research to encompass a broader range of material systems and fabrication techniques, highlighting the enormous potential these advancements hold for the creation of diverse semiconductor quantum technologies.

Scanning tunnelling microscopy (STM) and atomic force microscopy (AFM) are the only techniques capable of imaging and manipulating single atoms with atomic-scale precision. Constructing stable, robust, and functional atomic-scale devices in semiconductors requires atomic manipulation within a strongly covalently bonded system, where atoms typically cannot be repositioned but must be precisely placed on first contact with the surface. In the early 1990s, pioneers in this field developed a method to remove individual hydrogen atoms from a hydrogen-terminated silicon surface, known as hydrogen depassivation lithography (HDL). HDL enables the creation of chemically reactive sites down to a single dangling bond on a single silicon surface atom. The historical development and the current and future challenges for HDL are presented in Sections 2.2 and 5.4, while Section 3.4 presents an overview of the potential to realize spin and charge quantum devices based directly on silicon dangling bonds. Despite the astonishing level of control, including the ability to erase a dangling bond by the controlled re-deposition of a single hydrogen atom, large-scale system development faces key challenges. These include improving write speed, throughput, tip positioning fidelity, and tip reliability.

HDL remains the only tool capable of placing dopant atoms in silicon with atomic-scale precision. This is done by spatially controlling the chemical reactions of precursor molecules like phosphine

(PH$_3$) on the silicon surface. The deterministic placement of individual phosphorus atoms was first achieved in 2003. This breakthrough, combined with fiducial markers for device localization, homoepitaxial silicon overgrowth, and precise electrical contact fabrication, has enabled the creation and measurement of various groundbreaking quantum electronic devices forming a toolbox of critical components for future integrated quantum circuitry. These include atomic-scale wires, tunnel junctions, charge sensors, single-electron transistors (including the single atom transistor), arrayed few-donor devices, and one- and two-qubit gates. Three-dimensional devices have also been demonstrated using successive patterning and epitaxial growth stages. This remarkable progress in phosphorus-in-silicon-based quantum device fabrication and measurement is described in Sections 2.3 and 3.1. Challenges include precisely controlling the number and positions of phosphorus atoms, managing the surrounding crystal environment and other factors that influence the donor wavefunction, and reducing or eliminating sources of charge noise. Methods for improving the phosphorus incorporation accuracy and reliability, including STM-tip stimulated phosphorus atom incorporation and machine learning tools for accuracy and reproducibility, are discussed in Section 4.2.

The implantation of dopant atoms in semiconductors using ion beam systems dates to the very beginnings of the semiconductor industry. Since then, significant efforts have been made to develop highly focused ion beam systems capable of delivering individual dopants at precise locations for the fabrication of atomic-scale devices. Developed in parallel to the STM-based approach, single ion implantation has been an important component to the field. The first single-shot spin readout, electron spin qubit, and nuclear spin qubit in silicon were achieved using ion implantation. The developments, prospects, and fundamental challenges in this area are detailed in Sections 2.4 and 5.1. Two principal challenges include increasing the implantation confidence for single ions in the keV energy range and reducing the spatial straggle in both the device plane and the implantation depth. While sub-100 nm spatial precision is routine, achieving the positional accuracies of approximately 15 nm required for device architectures such as the Kane quantum computer remains a challenge. However, this limitation may be negated by the use of hybrid architectures where qubit entanglement is mediated by optical cavities, photonic waveguides, or superconducting circuits. Single ion implantation has the potential to operate at an industrial scale where millions of qubits are required for large-scale systems but is also versatile with a variety of dopant and impurity species possible.

Phosphorus in silicon has been the dominant material combination since the inception of the field, but the range of impurities that could be studied with this technique is large offering a whole new frontier of possibilities. A phosphorus impurity in silicon already provides a 4-dimensional Hilbert space (2 electron × 2 nuclear spin states) for quantum information processing compared to just two with a single spin qubit. However, as we progress down column V of the periodic table, we see that the $I=7/2$ nuclear spin of antimony can offer a 16-dimensional Hilbert space. Such a high spin nuclei has the potential to encode an error-corrected logical qubit within a single substitutional impurity, which is discussed in Section 3.2. It is also interesting to consider donor and acceptor atom qubits. Whilst donors in silicon exhibit multi-valley conduction band effects, acceptors are dominated by strong spin-orbit coupling. Thus, donors and acceptors each present unique advantages and challenges, suited to distinct architectures like all-electrical control for acceptors. Section 2.1 provides an overview of the fundamentals of these defect states, the gate-based quantum devices and analogue quantum simulators that can be built from them, and how these are modelled with theoretical and computational methods. The same properties that give single-impurity qubit arrays their strength in quantum computing and simulation also make them difficult to model accurately. Close interaction between theory and experiment, along with advancements in theoretical methodologies, will enhance our understanding and development of large-scale quantum technologies. A powerful experimental technique allowing detailed comparison with theoretical predictions is the direct measurement of bulk dopants using STM and scanning tunnelling spectroscopy (STS), as presented in Section 3.3. Although typically a surface technique, STM/STS can directly image dopant atoms near the surface (~< 10 nm) due to the large extent of the dopant wavefunction. Individual dopants in active devices have been examined this way, and future access to their spin degree of freedom is anticipated.

Developing methodologies for positioning impurity species beyond phosphorus using HDL requires significant effort to understand the surface physical and chemical processes of dopant precursor compounds. One now well-understood system is the incorporation of arsenic into silicon using arsine ($AsH_3$) as a precursor, as detailed in Section 4.1. Arsine dissociates more rapidly than phosphine exhibiting a ~97% success rate of dopant incorporation. Transitioning from phosphine to arsine may provide an alternative path to creating large-scale donor qubit arrays. Arsine has also been studied on the germanium surface, where arsenic atoms incorporate at room temperature without requiring a thermal incorporation anneal. Combined with other desirable characteristics, such as larger Bohr radii, stronger Stark effect, with a reduced sensitivity to exchange coupling oscillations, this makes germanium, and particularly arsenic donors in germanium another system with exciting potential for atomic-scale dopant devices and scalability.

For many other impurity species of interest for quantum applications, simple hydride analogues of phosphine are not stable, necessitating the search for new precursor compounds. Success in this area has so far concentrated on halide compounds such as boron trichloride ($BCl_3$) and bismuth trichloride ($BiCl_3$) for introducing acceptor and deep-level donor species, and phosphorus tribromide ($PBr_3$) for enhancing our understanding via chemical analogy. The search for other precursors may yet enable new functionalities in the fabrication process, such as semi-automated vertical alignment for three-dimensional devices. This, and other aspects of precursor development is discussed in Section 4.3. Moreover, many exciting opportunities exist for technologies beyond quantum computing: Section 5.6 presents an overview of potential applications in optoelectronics and quantum optics, energy-efficient digital electronics, analogue sensors, and superconductivity that may be enabled by the development of new precursors, including those for optically active impurities like erbium. Key challenges here include understanding the surface chemistry of new precursors, requiring the concerted effort of both STM experimentalists and computational chemists, and demonstrating compatibility with spatial control via HDL. To that end, another active area of research is the development of halogen resists and the extension of HDL to halogen desorption lithography (conveniently retaining the same acronym), as described in Section 4.4.

HDL, ion implantation, and the electrical measurement of laboratory-scale devices are the backbone of the field. However, significant efforts are now underway with disruptive potential. Progress has been made towards photon-based HDL using extreme ultraviolet (EUV) light to increase throughput. Hydrogen desorption using EUV has been achieved, as has the direct patterning of silicon using interference lithography, as discussed in Section 5.3. This technique may be used in conjunction with STM-based patterning where atomic-precision is needed, or as a stand-alone tool where multi-donor dots and interconnects are required. A key challenge is maintaining substrate cleanliness and demonstrating dopant incorporation in sub-100 nm patterns. Manipulating dopants in bulk using electron beams is another promising area. Combined with advances in machine learning control systems, individual dopants in three-dimensional bulk semiconductors can now be manipulated by electron beams, as discussed in Section 5.2. While understanding electron beam effects and developing machine learning methods is ongoing, this technique could dramatically impact future fabrication and enable correction of dopant positioning errors, for which no technology currently exists. Another possibility is atomic-scale fabrication of 3D components such as electromechanical systems, opening new possibilities for semiconductor devices, as described in Section 5.5.

This roadmap outlines the significant strides made in the development of atomic-scale semiconductor devices, from the exact precision placement of dopants using scanning probe techniques to the ~15 nm placement using ion implantation and the exploration of new impurity species and advanced fabrication techniques. These innovations are paving the way for scalable quantum computing and other cutting-edge technologies. By addressing current challenges and leveraging emerging methodologies, we are on the cusp of realising the full potential of atomically precise fabrication. The sections that follow provide a detailed look at these advancements and future directions.

# Section 2.1 – Shallow dopants in silicon; theory

Andrew J. Fisher[1] and Eran Ginossar[2]

[1]London Centre for Nanotechnology and UCL Department of Physics and Astronomy, University College London, UK
[2]School of Mathematics and Physics, University of Surrey, UK

**Status**

*Fundamentals*

The theory of isolated shallow impurities goes back to effective-mass theory, where the ground state of an electron bound to an isolated donor in silicon is the $1sA_1$ multivalley coupled wave function constructed from states near the six conduction-band minima (see Figure 1). Modern developments include atomistic theory, often based on tight-binding approximations, [Salfi2002, Martins2005], and explicit inclusion of multi-valley effects based on first-principles calculations [Gamble2015]. Acceptor states form from a single valence-band maximum at the zone centre but are dominated by strong spin-orbit interactions within these p-like states. However, because of the long range of both the Coulomb potential and the wavefunctions, it remains very challenging to describe either donors or acceptors using fully first-principles methods such as density functional theory.

*Quantum computing*

Electrons on different donors can tunnel between impurities as well as repel each other, giving rise to an effective exchange interaction between neighbouring spins [Gonzalez-Zalba2014]. Coupling can also occur via nuclear-electron spin state and electric dipole interactions and via superconducting resonators [Tosi2014].

Such coupled impurities have been long proposed as candidates for quantum processors. The nuclear or electron spin states are typically chosen to hold quantum information and can be operated on and measured with high fidelity. Entanglement of donor qubits is achieved via controlling the exchange interaction [Mohiyaddin2017]. The nuclear spin has especially long coherence times measured in many milliseconds, compared to typical 100microseconds for superconducting qubits. Initially, readout involved converting the spin state to a dc current; recently, much faster readout based on RF techniques has been developed. Fast (below 1ns) two-qubit $\sqrt{SWAP}$ gates based on electrostatic control of the exchange interaction between multi-donor quantum dots were experimentally demonstrated by [He2019], with an estimated fidelity of 86.7% (compared to 99-99.9% typical two-qubit fidelities with transmon qubits).

*Quantum simulation*

Arrays of donors are candidates for analogue quantum simulation given the relatively high degree fabrication and control [Salfi2016]. Their description by the Fermi-Hubbard model with disorder and long-range interactions has relevance to topological insulators, the metal-insulator transition and high-temperature superconductivity. On the insulating side it is plausible that the electron gas can be cooled to below certain many-body gapped phases through its phonon bath. By controlling the concentration of the donors and its chemical potential this system can behave as an insulator or metal and can be probed via transport measurements [Wang2022, Le2017], see Figure 2, extending to the fabrication and probing of donor topological insulators [Le2020, Kycinski2022].

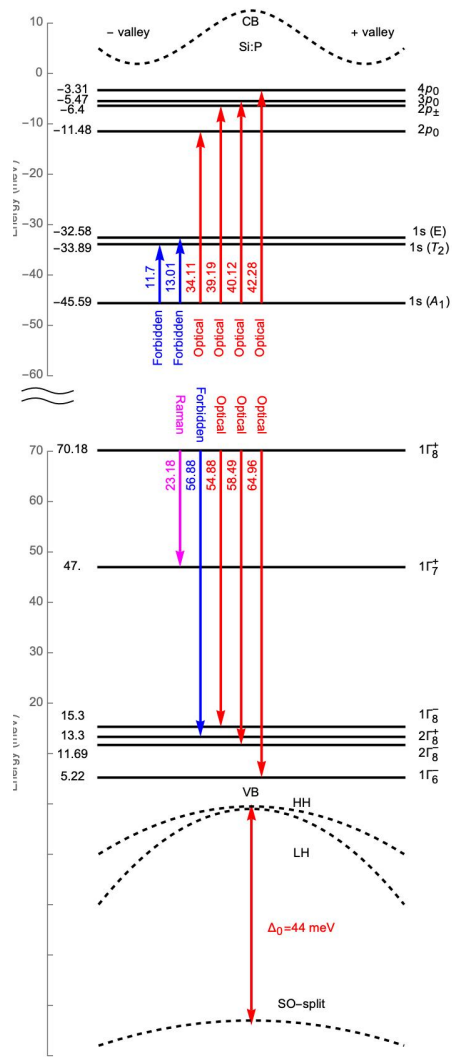

**Figure 1** Schematic diagram of the relevant energy levels for shallow donor (top part) and acceptor (bottom part) impurities in Si. The zeros in the energy scales are taken at the conduction and valence band edges respectively; the positions of the impurity levels are to scale but the shapes of the bulk band-structures are schematic only. Note the existence of conduction-band valleys displaced from the Brillouin zone centre and the significant spin-orbit splitting in the valence band. Where indicated, the levels are taken from experimentally observed transitions; otherwise, they are from calculations.

**Current and Future Challenges**

*Quantum computing*
Impurities in silicon have naturally long spin coherence times (thanks to low nuclear spin densities), high qubit density and compatibility with existing fabrication routes. However, the oscillatory structure of the multi-valley donor wave function, coupled with positional variability, creates variability in the exchange interaction which is a challenge for realising high-fidelity two-qubit gates with the natural spin qubit encoding. The short range of the exchange interaction also requires close spacings, providing a challenge for individual addressing of qubits, especially if one-qubit operations are implemented using magnetic fields. Qubit encodings compatible with electric field manipulation, including singlet-triplet encodings with two electrons or hole-based qubits, would be advantageous;

however, controlled implantation of acceptors is not yet developed. Forming isolated single donors is also a challenge, with multiple-donor quantum dots forming instead; modelling their variability remains challenging.

While single qubit gates and readout have been demonstrated with high fidelity, two-qubit gates remain challenging. Electric field noise arises from both local impurities and from nearby gates; this noise is correlated between neighbouring qubits because of the very small separations, leading to correlated errors which are challenging to correct. The long-term goal of most quantum computing efforts, a fault-tolerant processor, requires error correction. The most popular codes are intrinsically two-dimensional topological codes such as the surface code; while these are robust and map well to the structure of a 2D implanted layer, requiring only local interactions, they have very large qubit overheads. Recent attention has focused on quantum low-density parity check (LDPC) codes which are much more efficient but require non-local interactions.

*Quantum simulation*
Theoretical treatments must include both disorder and long-range interactions. The range of interactions depends on intrinsic factors such as density but also on screening effects by nearby electrodes and these need detailed modelling. Acceptors bring in spin-orbit interaction with added possibilities for engineering topological insulator phases of dopants, but the presence of interactions complicates the classification of such phases. In addition, new probes are required for large insulating arrays where conductance spectroscopy is not feasible on the insulating side.

*Underlying theory*
It remains a challenge to reconcile the extended nature of donor and acceptor states with first-principles theory. Furthermore, while the natural exchange interactions between donors are relatively well understood, the much richer set of couplings possible for acceptors (exploiting the spin-orbit coupling present in the valence band) are only just starting to be explored.

**Advances in Science and Technology to Meet Challenges**
*Quantum computing*
Improved understanding of noise sources will be required, as well as of efficient quantum codes that might be suitable for dopant arrays. In turn these are likely to require increased effort to engineer longer-range interactions, either by shuttling electrons between donors/dots [Mills2019] (as recently demonstrated with neutral-atom systems and ion traps) or by coupling them to microwave cavities [Morello2020](as is routinely done with superconducting qubits). Additionally, since the dopant arrays are strongly coupled, they may be candidates for novel approaches such as *quantum reservoir computing*, though more work is required into encoding and decoding information from the array [Sutherland2024].

*Quantum simulation*
Exact diagonalisation studies of the extended 2D Fermi-Hubbard model are limited to very small arrays typically up to 5x5 impurities. Exploring larger arrays is essential to removing finite size effects and estimating the properties of gapped phases. Numerical approaches which involve variational Quantum Monte Carlo or Tensor Network techniques have been able recently to target the local model in the underdoped regime. Further expected development in these techniques will enable their application to the disordered Fermi-Hubbard model with extended interactions and disorder, spin-orbit interactions and multi-dopant arrays which are present in the dopant array system. Such studies are required for gaining a better understanding about the low temperature phase diagram and whether experimentally the system can be designed to simulate high-temperature superconductivity, quantum spin liquids or other exotic phases. It will be important to quantify the deviations from approximations made in quantum simulations and maybe also broadening the range of models beyond Fermi-Hubbard.

*Underlying theory*

Improvements in the understanding of large-scale dopant structures (including multi-dopant arrays and multi-dopant dots) are required. These could exploit efficient tensor-network approaches as well as local approaches to the electronic structure. The field also requires a better understanding of the probes of relevant excitations (especially spin excitations, the most relevant to typical qubit applications) in transport measurements; it could also be valuable to combine transport with other probes, such as optical excitation or spin resonance. Finally, improved theory is required on the atomistic scale to understand new growth and implantation processes (e.g. to produce deterministically implanted acceptors).

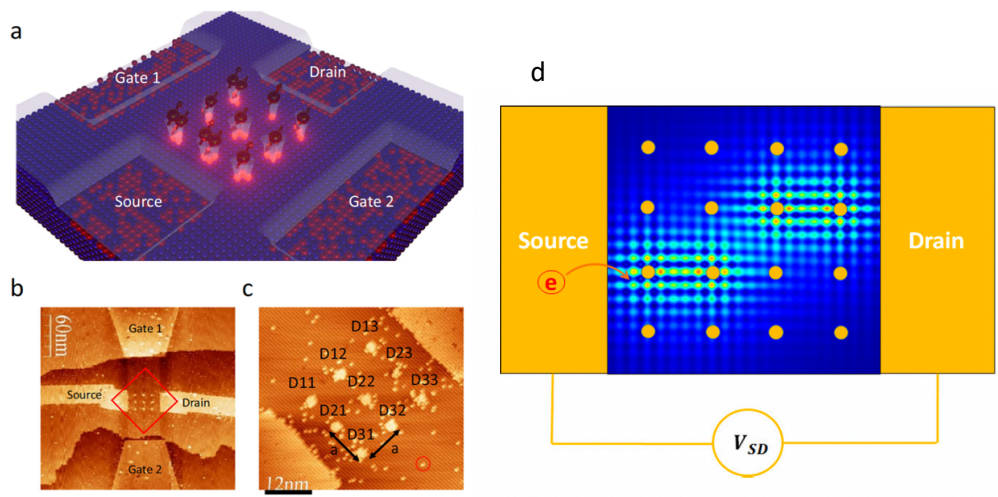

**Figure 2** 2D dopant-based quantum dot arrays as a platform for simulating the extended Fermi-Hubbard model (a) Schematic of the experimental Fermi-Hubbard system composed of a 3 × 3 array of single/few-dopant quantum dots coupled to in-plane gates and source-drain leads, allowing transport measurements through the array. (b) STM image of the central device region of the 3 × 3 array acquired immediately following hydrogen lithography. (c) Atomic resolution STM image of the 3 × 3 array pattern (zoom in of marked square region in b) [Wang2022] (d) A 2D donor array coupled to a source and a drain under a bias $V_{SD}$. The chemical potential of the leads can be varied by a gate voltage. Electrons from the source can tunnel through a many-body state of the array that is delocalized along a path that connects one side of the array to the other.

**Concluding Remarks**

The same features (their small size and integration with the solid-state environment) that make arrays of single-impurity qubits powerful for both quantum computing and quantum simulation, also make them challenging to model. Improvements in understanding will require both developments in the theoretical methodologies and close interaction between theory and experiment.

# Section 2.2 – Hydrogen Depassivation Lithography

Joseph W. Lyding

Department of Electrical and Computer Engineering, University of Illinois, Urbana, Illlinois 61801, USA

**Status**

Hydrogen depassivation lithography (HDL)[1] was developed with the intent of creating an atomic resolution STM-based lithography tool for technologically relevant silicon surfaces with patterning that is robust at room temperature. Inspiration for HDL came from Eigler's writing of 'IBM' with 35 xenon atoms on a nickel surface[2]. Before HDL development started at Illinois, efforts were underway at NIST, Bell Laboratories, and IBM that could have led to HDL. Dagata at NIST used an ambient STM to create local oxidation patterns on a wet-passivated Si(100) surface[3]. Becker at Bell Labs showed that electron stimulated desorption of hydrogen from a wet-passivated Si(111) surface created regions of 2x1 reconstructed clean silicon[4]. Boland at IBM showed that ideal Si(100)2x1 monohydride surfaces could be created in his studies of recombinative thermal desorption of hydrogen[5]. The original HDL study showed two hydrogen desorption regimes; one at higher electron energies that is only electron dose dependent, and higher resolution mode at lower electron energies that is electron flux dependent[1]. A follow-up detailed study showed that the low-energy high-resolution patterning mode is consistent with vibrational heating of the Si-H bond[6]. Feedback controlled lithography (FCL) was developed to facilitate deterministic single H atom patterning to create precision templates for local chemical modification such as single molecule adsorption[7]. In addition to this chemical contrast, HDL patterns exhibit electronic contrast with single atom wide patterns being metallic and dimer-wide patterns being semiconducting[8]. Figure 1 illustrates this with an HDL line that transitions from single atom wide metallic to dimer wide semiconducting behaviour.

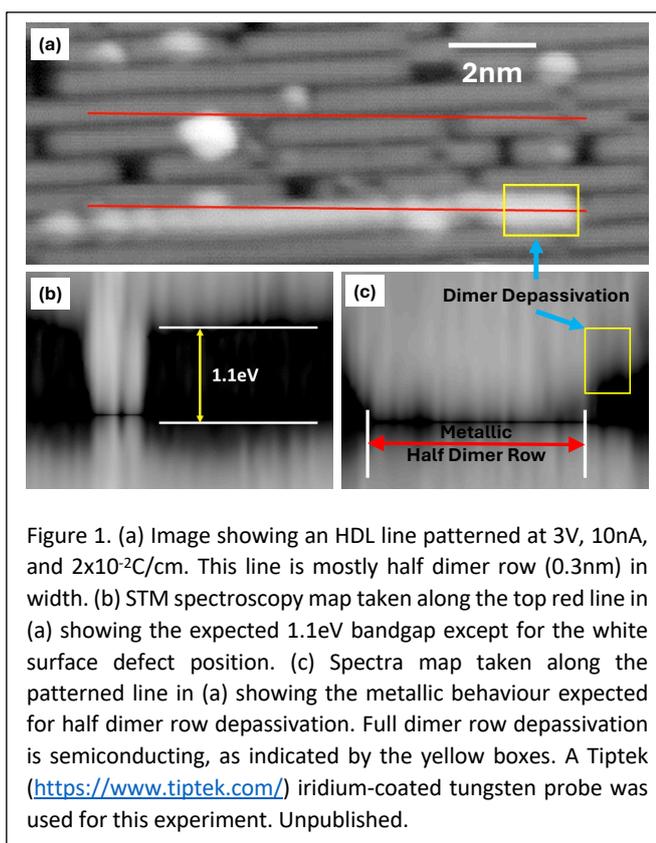

Figure 1. (a) Image showing an HDL line patterned at 3V, 10nA, and 2x10$^{-2}$C/cm. This line is mostly half dimer row (0.3nm) in width. (b) STM spectroscopy map taken along the top red line in (a) showing the expected 1.1eV bandgap except for the white surface defect position. (c) Spectra map taken along the patterned line in (a) showing the metallic behaviour expected for half dimer row depassivation. Full dimer row depassivation is semiconducting, as indicated by the yellow boxes. A Tiptek (https://www.tiptek.com/) iridium-coated tungsten probe was used for this experiment. Unpublished.

In the 30 years since its inception, HDL remains the only atomic resolution fabrication tool available for silicon surfaces. The extreme chemical contrast between the patterned and non-patterned areas affords the opportunity to perform chemistry with atomic site precision. As such, HDL is the core nanofabrication engine for the worldwide scanned probe based silicon quantum computing effort[9]. Here, HDL creates the adsorption sites for single phosphine molecules that ultimately incorporate the single phosphorus atom qubit cores, as well as the in-plane degenerately phosphorus δ-doped metallic wires with atomically precise edge definition. An unexpected byproduct of HDL was the discovery of a giant isotope effect for deuterium desorption from Si(100) surfaces[10]. This led to the idea[11] of using deuterium to retard hot-carrier degradation in CMOS technology, which is currently being practiced by several major chip producers worldwide.

**Current and Future Challenges**
While HDL is an atomic resolution technique it is saddled with the throughput and positioning fidelity of scanned probe piezoelectric positioners. However, for proof-of-concept demonstrations and technologies such as silicon quantum computing where a few qubits confer a lot of quantum computing capability, HDL is adequate in its current form. Generally, atomic resolution HDL can be done under a wide range of conditions. A typical set of patterning conditions for FCL or atom-wide line writing might be 3V and 10 nA (see Figure 1), with a write speed of 10 nm/s corresponding to an electron dose of $10^{-2}$ C/cm. Increasing the current to hundreds of nA, and reducing the voltage to 2 V retains the patterning resolution while increasing the writing speed to 1000 nm/s. One can draw a simple comparison with state-of-the-art extreme UV (EUV) optical lithography by estimating the equivalent probe velocity for HDL to operate at EUV speeds. Assuming, for example, that EUV can write a square centimetre die with lines on a 10 nm pitch and that 100 die can be written per second, corresponds to linear write speed of $10^{14}$ nm/s, or eleven orders of magnitude faster than HDL.

Probe issues are also a challenge for HDL. The most common probes are made of tungsten and are prone to oxidation, tip changes due to tungsten atom motion in the high-field apex region, and durability against the unintended instances of strong tip-surface interaction. Another issue is consistent probe-to-probe functionality. Differences in probe apex geometry affect the electric field distribution, current density, and electron beam diameter of the probe. While techniques like FCL can help compensate for these differences, open-loop patterning would yield high probe-to-probe inconsistencies.

**Advances in Science and Technology to Meet Challenges**
Advances in probe technology are needed to make HDL robust for long writes that require atomic resolution for key features and edge definition, such as in silicon qubit fabrication.[9] Ultra-hard probe coatings, such as with the metallic ceramic $HfB_2$, and sharpening techniques like field-directed sputter sharpening (FDSS) are steps in this direction[12]. Minimizing unwanted probe-surface interactions are needed for consistent probe performance and better probe longevity. An important step in this direction is to stabilize the STM feedback loop by actively tuning it to compensate for changes in the local tunnelling barrier height that are present for even the best prepared clean surfaces[13]. To address the throughput issue, MEMS based scanners with the potential of scaling to $10^7$ scanners over the area of a 300 mm wafer are under development[14]. Though very impressive, the total throughput will still be orders of magnitude too slow for CMOS chip processing. However, one can envision using STM and HDL in much the same way that chip manufacturers use electron beam lithography for prototyping and mask generation. A potential target for HDL would be in the generation of masks for nanoimprint lithography (NIL). In this case, selective chemistry following HDL could produce an etch mask for a directional reactive ion etch (RIE) process to transfer the pattern into the silicon substrate with minimal loss of fidelity. This mask could be used as is for NIL, or it could be coated using an atomic layer deposition (ALD) process to yield a mask surface that has better properties for NIL.

**Concluding Remarks**
In the 30 years since its invention HDL[1] remains the only technique capable of patterning a hydrogen resist on Si(100) with atomic precision. HDL works at covalent bond energies and therefore produces patterns that are stable at room temperature. The extreme chemical contrast between the patterned and non-patterned areas enables local chemical modifications including molecular adsorption with single molecule resolution[7, 9, 15]. The giant deuterium isotope effect discovered in HDL experiments[10] led to the current use of deuterium processing in CMOS fabrication to retard hot-carrier degradation effects[11].

**Acknowledgements**
The development of HDL was funded by the Office of Naval Research under Grant N00014-92-J-1519.

## Section 2.3 – P in Si device fabrication

Richard Silver, Fan Fei, Pradeep Namboodiri and Jonathan Wyrick

Atom Scale Device Group, National Institute of Standards and Technology, Gaithersburg, MD 20899, USA

**Status**

P dopant placement using hydrogen-based scanning probe depassivation lithography has enabled the creation of atomically engineered devices and materials whose properties can be designed based on the exact position and number of dopant atoms. Using atomic precision P-dopant placement and encapsulation in an epitaxial silicon layer, a variety of functional devices have been fabricated, including single atom transistors [1,2], donor/quantum dot devices [3], and arrayed few-donor devices for analogue quantum simulation [4]. Complex atomic arrangements of single or groups of atoms are possible, and device performance can be tailored by the specific atomic configurations and edge geometries in a single electron transistor (SET) charge sensor, qubit, or lattice site. Since devices directly depend on individual atomic species and their interactions, they are sensitive to atomic imperfections, defects, and precise dopant positions, necessitating development of robust lithographic processes and epitaxial silicon overgrowth to ensure a pristine crystalline environment free from defects while using temperatures low enough to avoid inducing atomic movement.

The complete process is complex and typically includes pre-patterned fiducial marks for device re-location, Si surface reconstruction, H passivation, STM patterning, dopant precursor exposure and incorporation, and silicon overgrowth, followed by ex-situ steps to locate the buried devices and subsequent cleanroom processes to make contacts or place top gates and electron spin resonance (ESR) lines, see Figure 1 [5]. These additional ex-situ process steps must be compatible with the Si patterning, P incorporation and overgrowth. However, once a device is encapsulated in epitaxial silicon and external processes are completed, devices can be stable for years. Contact to buried atomic structures using silicide processes allows repeatable contact to complex devices [6].

Reliable device performance is highly dependent on defining stable atomic dimensions for tunnel coupling between source, drain, qubits, array sites, and other device components. Tunnel rates can be controlled by patterning with atomic precision at the level of single dimer rows and exquisite control of overgrowth and process parameters has yielded well-behaved tunnel junctions at the atomic scale [7]. Future engineering of atomic-scale properties can be enhanced by combining specific dopant configurations and their alignment with substrate axes to further engineer the electron wavefunction and hopping properties.

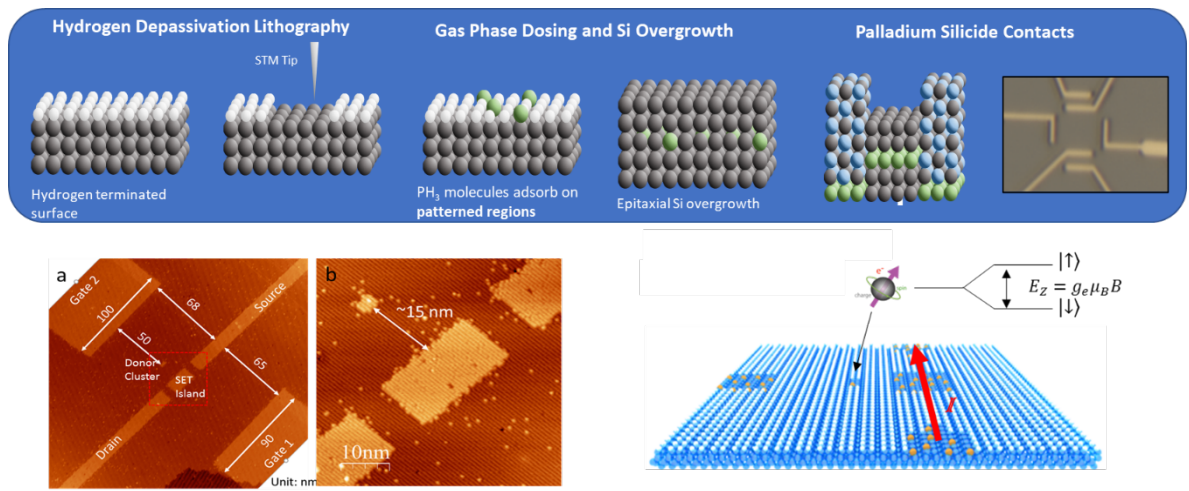

**Figure 1** Upper schematic shows the basic fabrication steps from hydrogen termination to patterning, overgrowth and contacts. The lower figures show a donor/quantum dot device schematic (right) and an actual device at the patterning stage (left).

**Current and Future Challenges**
While there are several challenges to robust device fabrication, the most critical are maintaining the number and positions of dopant atoms throughout the fabrication process and achieving overgrowth that results in low noise devices and large gate ranges. The homogeneity of atoms in small structures such as qubits as well as larger single electron charge sensors and their leads plays a role in device functionality. Advances in atomic positioning and better control of configuration geometries will enable improved device yield and more robust device performance.

Fabrication enabling three-dimensional device operation has been demonstrated [8] and is likely a requirement in scalable qubit architectures. This can be achieved by extending STM device fabrication to the vertical direction and adding well-aligned top gates. The challenge is re-establishing atomically ordered and monolayer flat surfaces for subsequent atom-scale patterning.

One ongoing process challenge is that maintaining low temperatures during silicon overgrowth, yet achieving high quality epitaxy, mandates low growth rates in the monolayer per minute range, see Figure 2. The sensitivity to crystalline overgrowth defects, individual contamination atoms/molecules, and charge traps motivates fabrication in an exceptional UHV environment with maximum pressures in the few $10^{-11}$ torr (~$10^{-9}$ Pa) range.

Metrology presents unique challenges due to the need for sensitivity at the single atom level while having to make many measurements at low temperature on actual quantum devices. We need to quantify atomic positioning and subsequent movement during fabrication, identify individual atomic contaminants, and characterize epitaxial quality to ensure pristine, crystalline encapsulation [9]. It is essential to understand the effects of subtle variation in atomic fabrication processes on device characteristics such as coherence times or tunnel coupling. Monitoring dimensions post process usually requires low temperature electrical metrology of load/unload tunnel rates, SET on-state currents, and charging energies. Weak localization and coherent electron scattering can be used to characterize delta layer electronic quality and help understand atomic impurity scattering processes. Transport methods or RF reflectometry capable of measuring atom-scale imperfections are often the best solution because conventional transmission electron microscopy (TEM) and atom probe tomography have limited utility due to averaging and their destructive low throughput attributes.

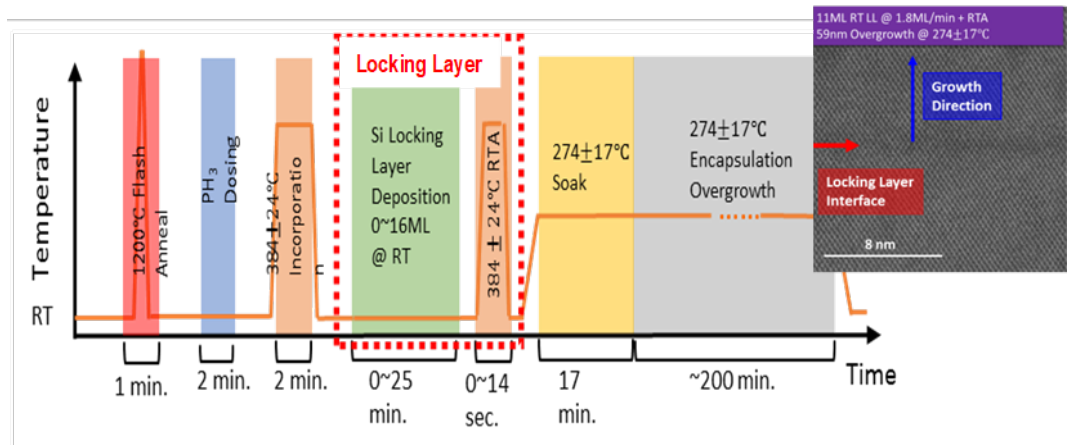

**Figure 2** An example of the in-situ dosing, incorporation, and overgrowth steps. While the TEM shows good confinement of a delta layer (red arrow), detailed atomic information is better gleaned using low temperature electrical measurements.

**Advances in Science and Technology to Meet Challenges**

One goal of atomically precise fabrication is building large homogenous arrays of atoms to realize the analogue quantum simulation (AQS) of the extended Hubbard model in 2D. Likewise, fabricating atomically engineered materials or arrays of qubits whose properties depend on the detailed atomic configurations requires long range homogenous, rectilinear device fabrication. The STM, as the core fabrication tool, needs to undergo further improvement in linearity and motion control, while the STM tip remains an area of artform, lacking quantitative repeatability at the atomic scale. The atomic geometry of the tip has a direct effect on lithography performance and can add significant variability to lithography performance and longevity. This is an ongoing challenge with STM for routine imaging and even more so with lithography at the atomic scale due to additional performance constraints on tip geometry and repeatability. Field ion field electron microscopy (FIFEM), SEM, and TEM can be used to improve STM tip performance through selectivity or direct shaping of the tip and apex geometry.

Reliable models that allow us to understand the detailed band structure and variations in densities of states in device components help to define fabrication tolerances and drive new design geometries. Developing a comprehensive physics-based understanding of single and few atom "artificial molecules" that form the basis of qubits and sites in a quantum simulation array will facilitate the design and fabrication of specific atomic geometries for qubits and array sites [10]. Tunnelling rates, spin filling, and electron densities are all determined by details of the inter-atom hopping rates, Coulomb interactions, and magnetic interactions. Simulation tools such as DFT and tight binding methods are increasingly essential to map out the energy spectra and magnetic ordering in donor/quantum dot devices or sites in a Hubbard array.

Different dopant atoms have the possibility to expand the accessible physics that can be explored and potentially reduce some of the fabrication challenges. Introducing new spin manifolds using alternative dopants offers new directions in quantum device design. Recent work exploring new methods for dopant incorporation and activation will likely improve incorporation probabilities and have a significant impact on long term progress.

**Concluding Remarks**

Precise placement of atoms in the solid state offers an unmatched approach to designing and fabricating quantum devices and new quantum materials where the properties of such a device or material depend on the type and position of every atom. This sophisticated flexibility to engineer at the atomic scale is subject to the influence of individual atomic defects, imperfections, and broken bonds. Scalable architectures have been envisioned and stable, long lifetime devices with good high frequency characteristics allowing coherent control over individual electrons have been demonstrated. However, this all depends on atomically precise fabrication and control of the crystalline environment in which the devices operate. There are several proposed research advances and new processes that will likely improve device functionality and fabrication repeatability. Further development of accurate simulation tools to comprehensively understand device performance at the atomic scale will provide increasingly important guidance to new device design and fabrication methods.

**Acknowledgements**
The author wishes to acknowledge useful discussions and technical contributions from Xiqiao Wang. This material is based in part upon work supported by the National Science Foundation under Grant No 2240377, and by the Department of Energy Advanced Manufacturing Office Award Number DE-EE0008311.

# Section 2.4 – Focussed ion beam as a tool to realize scalable one-million qubit devices

M.G. Masteghin, D.C. Cox, B.N. Murdin and S.K Clowes

Advanced Technology Institute, University of Surrey, GU2 7XH, UK

**Status**

For many decades, broad area ion implantation has been an industry standard in semiconductor engineering and fabrication, renowned for its precision and control in modifying semiconductor material properties [1]. The development of focused ion beam (FIB) systems in the early 1990s expanded the applications of ion beams to nanofabrication, including milling and ion beam-assisted deposition, and has become widely used in transmission electron microscopy (TEM) sample preparation. More recently, FIB has gained recognition for its potential in quantum technologies, particularly in the deterministic placement of single or few ions for impurity and defect engineering, with isotopic selectivity and achieving sub-100 nm positional precision [2]. A standard deterministic ion implanter system, illustrated in Figure 1, features a mass filter for selecting specific species and isotopes, along with a fast chopper for blanking the ion beam [3]. Significant effort is expended on the alternative quantum dot platform to produce a deterministic spectrum (which is usually electrical-gate-controlled) [4], and this is achieved naturally with implanted single ions (ISI) and defects. While FIB cannot achieve the precision required for Kane's impurity quantum computer proposal in which qubits are intricately coupled at separations of just a few nanometres via the exchange interaction, new schemes compatible with FIB capabilities have been proposed. One such scheme is the flip-flop qubit proposed by Tosi et al. [5], which enables long-range qubit coupling over distances of 100-500 nm via electric dipole-dipole interactions. This approach accommodates the vertical positional uncertainty of implanted impurities by tuning individual qubits to clock-transitions with multiple surface gates. Other innovations include the creation of extremely "quiet" host materials free from unwanted isotopes with non-zero nuclear spin, all achievable using focused ion beam techniques [6].

The versatility of FIB for defect and impurity engineering in creating colour centres is also highly significant for quantum technologies. These colour centres, such as the nitrogen vacancy (NV) centres in diamond, have a wide range of applications, including quantum computing, quantum sensing, and single-photon sources. Applications for single photon emitters require high reproducibility in large arrays [7] and again ISI offers advantages over dots. FIB is ideally suited for this task due to its ability to implant specific ion species and isotopes with high spatial accuracy in a small volume and (potentially) sub-Poissonian control of the number of ions implanted [2].

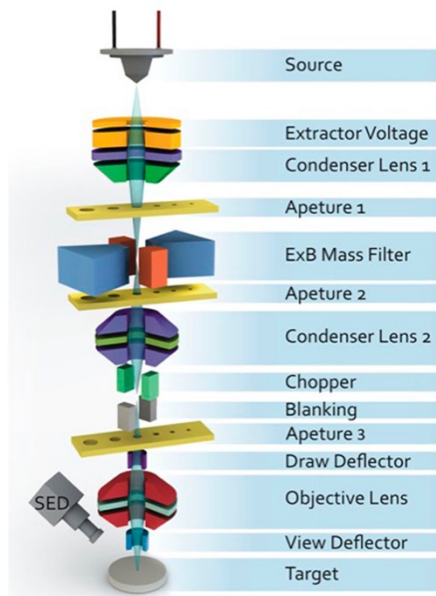

**Figure 2** Schematic of a typical deterministic ion implantation system. The ion beam is extracted from a liquid metal alloy ion or gas plasma source. The ExB mass filter and at the chopper-blanking regions defines the mass and the beam chopping resolution, respectively. The objective lens focuses the beam on the target plane. The draw deflector allows patterning of the implanted ions and secondary electron detection (SED) images of registration marks and detection of ion implantation events. *[3]*

**Current and Future Challenges**

**Variety of Substrates & Ion Sources:** Despite the known requirement for single impurities being based almost exclusively around group-III, -IV, and -V elements of the periodic table, as well as the lanthanide series, quantum technology platforms are very diverse. Therefore, a versatile FIB for ion implantation will only be realized if coupled with ion-source development that can manufacture optimized filament tip geometries and synthesize different alloys. Most FIB ion sources now required for quantum technology have been previously demonstrated in the research literature, but significant challenges exist in producing reliable, stable, and long-lived sources that provide practical utility. For instance, fabricating a million-qubit array using FIB, implanting at 10,000 singly placed ions per hour, would require a stable source over many hours. Additional complications arise as each species can present unique challenges in detection, activation, and scalability for the commercialization of methods, tools, and devices. It may also be required to implant multiple different species in the same device.

**Validation of Counting Accuracy:** FIB implantation through masks or apertures can be used to implant a plethora of ions with good positioning accuracy. However, without a robust mechanism to detect individual ion impacts, Poisson statistics limit the likelihood of a single implant to only 36%, with empty sites and sites with two or more ions making up the remaining 64% [2]. To achieve deterministic implantation, ions must be counted before or during impact. In the former, they can be extracted from an ion trap [8], and in the latter, generated electrons are sensed as secondary electrons [2] or ion beam-induced charge (IBIC) [9]. Each method has its advantages and disadvantages, balancing counting accuracy, speed, and flexibility of species.

**Assessment:** For those working on the development of the implantation, a major challenge is assessing the quality of the implanted sample. Single impurities or defects are, by their very nature, exceedingly difficult to measure. Therefore, metrology tools for sensitive, non-destructive post-imaging (pre- or post-annealing) of single- or few-atom impurities are needed, such as stimulated emission depletion (STED) microscopy [10] or nano x-ray fluorescence (XRF) [11].

**Advances in Science and Technology to Meet Challenges**

Future upgrades to existing XRF facilities might be the key to developing a large-area and high-resolution scanning methodology for single-atom detection. Due to the backscatter illumination and long probing depth of x-rays, XRF is not only free of any sample preparation requirements but also allows for the identification of single atoms beneath dielectric layers and/or metallic contacts (given they are contrasting elements). Required upgrades include the need for a brighter source and the reduction of background noise from lenses, detectors, and other beamline components. Unfortunately, the high running costs and limited availability will limit its widespread use in the semiconductor industry. In-situ damage healing and analysis would be tremendously beneficial.

The primary advancements must be aimed at improving existing FIB implantation tools. Even with a source that has great stability, low operating temperature, and low emission current—all contributing to the reduction of chromatic aberrations—it is still essential to have a Wien filter with high mass resolving power to distinguish species/isotopes with similar mass-to-charge ratios. A more challenging engineering endeavour would be the reduction of neutral generation due to scattering within the optics components. When neutrals are formed past a neutral filter (e.g., chicane-like lenses), the inability to focus these species jeopardizes the spatial accuracy of the implantation, even though detection efficiency might remain unaffected. Nevertheless, even with the ultimate FIB implanter, spatial accuracy will still be limited by fundamental properties such as ion straggle (which scales directly with energy and inversely with mass) and mobility during annealing. Therefore, final improvements must come from materials science through the development of annealing recipes that result in limited atomic mobility or controllably drive the implants based on free energy and Fick's laws. Lastly, especially for research machines, FIBs must be versatile. For example, they should allow implantation of isotopically pure host material, a variety of qubit/colour centre ion species, and possibly also local electrodes around the ion.

**Concluding Remarks**

Focused Ion Beam (FIB) instruments are extensively used in the semiconductor industry for circuit repair and inspection. Recent advancements in the deterministic implantation of single impurities suggest that their applications could expand to the direct writing of nanoscale devices for quantum technologies and optoelectronics, such as the creation of single-photon emitters and qubits. This capability brings the achievement of highly addressable defect and impurity solid-state qubits closer to reality, potentially with a single tool.

**Acknowledgements**

*Authors acknowledge financial support from the Engineering and Physical Sciences Research Council (EPSRC) [Grant No. EP/X018989/1, EP/X015491/1, and Grant No. EP/W027070/1].*

# Section 3.1 – Atomic-Precision Engineering of Atom Qubits for Quantum Computing

Joris G. Keizer[1,2,3] and Michelle Y. Simmons[1,2,3]

[1]School of Physics, University of New South Wales Sydney, Australia
[2]Silicon Quantum Computing Pty Ltd, Sydney, Australia
[3]ARC Centre of Excellence for Quantum Computation and Communication Technology (CQC2T), Australia

**Status**
The ability to use a Scanning Tunnelling Microscope (STM) to pattern a hydrogen resist, in combination with phosphine dosing and molecular-beam epitaxy, has been instrumental in manufacturing electronic devices in silicon at the atomic-scale. The combination of the exceptionally long coherence times of phosphorus atom spin qubits, with the scalability of the silicon materials platform, positions atomically engineered devices at the very forefront of the development of a universal, error-corrected quantum computer. Critical components of quantum integrated circuits, such as atomic-scale interconnects, single-electron transistors, single-lead charge sensors, charge sensing tunnel gaps, single-atom transistors, single-qubit gates, and two-qubit gates, have been demonstrated at the atomic-scale. A key advantage of this atomic-precision approach over more traditional top-down gate-defined approaches is that all the functional components are fabricated inside an isotopically pure, and high-quality defect free epitaxial silicon crystal, away from any noise inducing interfaces. This low noise environment, demonstrating the lowest observed charge noise in a qubit to date [KRA2020], lengthens the qubit coherence times and, as such, reduces the computational overhead needed for error-correcting schemes in a quantum computer. Furthermore, this ultra-high vacuum technique allows for three-dimensional manufacturing [KOC2019], in which multiple device layers can be stacked and aligned on top of each other with nanometre precision, all within the same silicon crystal. This three-dimensional manufacturing capability facilitates the realisation of low-noise qubit gating and control along with the possibility of all-epitaxial ESR and MNR antennas, epitaxial noise shielding layers, and on-chip integrated logic. The ability to engineer multiple device layers within a single crystal provides a pathway to realise scalable error-corrected universal quantum computing architectures. This is particularly the case for surface code-based architectures where the need to control and read out qubits synchronously and in parallel requires the formation of a two-dimensional array of qubits with control electrodes above and below the qubit layer [HIL2015]. Having demonstrated the capability to manufacture the key components for an error-corrected and scalable universal quantum computer, the further development of atomic-precision engineering is now largely driven by the desire to increase qubit and gate performance toward logical qubit demonstration.

**Current and Future Challenges**
Kane's theoretical proposal for a single atom-based quantum computer in 1998 was revolutionary. In Sydney we have pioneered atomic-precision manufacturing to realise reproducible, high-quality atom qubits in silicon with fast, stable operation. Key to the success has been understanding the role of spin-orbit coupling. Originally believed to be weak for electrons bound to phosphorus atoms in bulk silicon and resulting in long spin lifetimes (~mins), this lifetime was found, surprisingly, to be much reduced (~secs) in device architectures where metal gates are nearby, and the spin is operated in an electric and magnetic field. With careful alignment of external electric and magnetic fields, we showed that bulk spin-relaxation times were recoverable [WEBER2018]. This study highlighted the importance of understanding key factors that affect the extent of the donor wavefunction, such as charging energies, valley-orbit splitting, Stark shift of the electron g-factor, nuclear hyperfine, and exchange coupling. Gaining control over each of these properties with atomic-precision engineering remains a key focus. We can easily engineer the wavefunction by adding more phosphorus atoms to the qubit, resulting in stronger electron confinement, and longer electron spin lifetimes (~30 sec in a 3P register)

[WATSON2017]. The separation of phosphorus atoms within the register provides another control parameter, providing access to the molecular qubit regime where electrons bound to phosphorus atoms separated by larger distances ~8nm exhibit excellent qubit properties, surpassing those of single phosphorus atom qubits [KRANZ2023A]. The next level of atomic-precision engineering is to control the angle of the P atom separation with respect to the silicon crystallographic axis. The strength of the exchange interaction, critical in exchange based two-qubit gates, can fluctuate due to valley interference. Placing the phosphorus atoms along specific crystallographic directions removes the valley interference resulting in well-defined exchange coupling [VOISIN2020]. Placement of the atoms along well-defined crystallographic directions also facilitates the engineering of large hyperfine Stark coefficients (~70MHz/MV m$^{-1}$) allowing us to operate high fidelity (~99.99%) qubits rapidly without affecting neighbouring qubits [JONES2023]. In terms of gate operations, extremely fast (0.8ns) two-qubit exchange or √SWAP gates have been demonstrated between two phosphorus multi-nuclear spin registers in natural silicon [HE2019]. A key challenge for all solid-state qubit platforms ahead is to limit charge noise which causes unwanted fluctuations in the exchange coupling between electron spins. One way of mitigating the effect of charge noise is by atomic engineering a large magnetic field gradient (>800 MHz) between the qubits with CNOT gate fidelities of 99.98% achievable for a 2P-3P system [KRANZ2023B].

**Advances in Science and Technology to Meet Challenges**

Manufacturing reproducible high fidelity >99.9% single- and two-qubit gates requires the ultimate control in atomic-precision engineering: control over the number phosphorus atoms in the register and lattice-site precision in the placement of each phosphorus atom. This is just out of reach of current atomic-precision manufacturing techniques. Recent work on STM tip-assisted incorporation has demonstrated deterministic incorporation of single phosphorus atoms with atomic-precision [WYRICK2022]. Whilst an impressive technological achievement, the complex and time-consuming nature of the process, and need for an exceptionally stable STM tip, practically limits its application to low temperature (<77K) lithography and few sites. Other approaches are currently under development that control the incorporation kinetics through precursor dosing pressures, times, and temperatures (150°C or 77K) [IVIE2021, SIMMONS2022] or the use of a different species of donor atom [STOCK2024]. Results are promising, but more work is needed to validate the process and integrate it into atomic-precision device manufacture. As outlined above, the ability to deterministically place phosphorus atoms at controlled positions in the silicon crystal allows for precision engineering of qubit properties. The parameter space, however, is large and given by the relative qubit position, number, and relative position of atoms within the qubits, crystallographic direction, and number of electrons on the qubit, see Figure 1. A theoretical approach utilising atomistic modelling to predict qubit properties and performance is therefore of high value. Historically, atomistic modelling has focused on developing an understanding of experimental observations. However, over time this has now reached the stage where it is becoming a predictive tool for qubit engineering [HSUEH2024]. The same is true for larger scale processor components such as charge sensors, tunnel gaps, electron reservoirs, and control gates. While these components have been successfully incorporated for many years in atomic-precision devices, and are experimentally well understood, their integration within large-scale architectures is still under development. Here, multi-scale modelling of full device structures [DONNELLY2023], see Figure 2, can assist in rapid development, not only for atom-based quantum computers but for novel transistor designs that leverage the ultra-sharp dopant confinement potentials offered in atom-scale devices.

**Concluding Remarks**

Atomically precise manufacturing of phosphorus atom processors has evolved from the first single-atom transistor in 2012 to the first integrated circuit made with atomic-precision in 2022 [KICZYNSKI2022] and now to multi-nuclear spin qubit registers in 2024 [REINIER2024]. The excellent properties of these qubit registers in isotopically pure silicon can be further engineered through exact atomic-precision placement of the phosphorus atoms in these registers. It has been shown that this level of engineering is indeed possible using tip-assisted incorporation, with scalable incorporation techniques currently under development. Manufacturing of the accompanying large-scale integrated control and sensing components is achievable for qubit counts to ~1000 with current processes. For larger qubit numbers, automated manufacturing will need to address tip stability, feedback control, lithography automatization, and possibly parallelization to ensure practical manufacturing times for large scale processors.

**Acknowledgements**
Since 2000 this work was undertaken in the Centre of Excellence for Quantum Computation and Communication Technology. Since 2017 the company Silicon Quantum Computing has been established to manufacture full-stack processors based on the excellent qubit properties achieved.

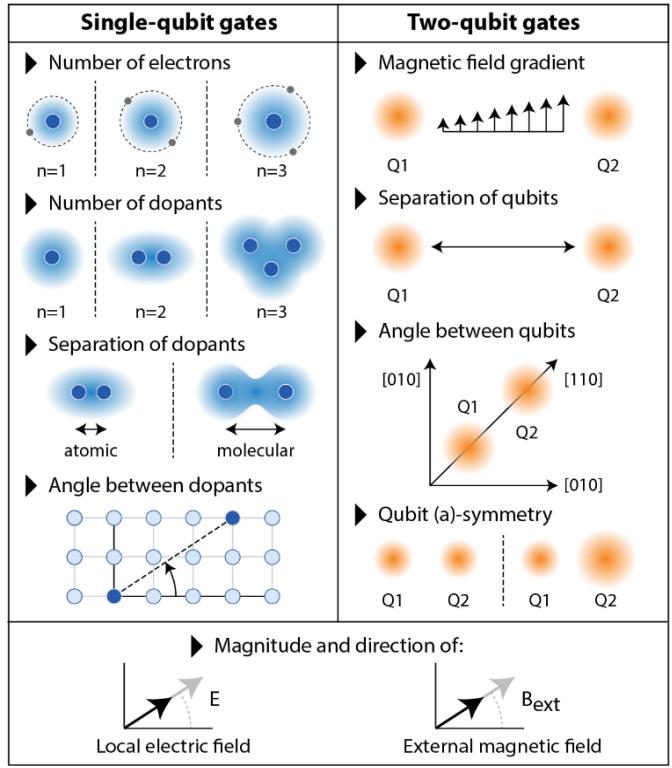

Figure 1: **Precision engineering of single- and two-qubit gates for atom qubits in silicon.** For single-qubits gates, both the number of electrons and number of phosphorus atoms determine the size of the electron wavefunction. Additionally, increasing the separation between phosphorus atoms allows the electron to transfer from an atomic state to a molecular state, affecting $T_1$ relaxation times and $T_2$ decoherence times. The angle between the phosphorus atoms relative to both the crystal axis and the local electric field enables engineering of both the hyperfine Stark coefficients and the spin-orbit interactions, which together with alignment to the external magnetic field can control the $T_1$ relaxation times, $T_2$ decoherence times and $T_2^*$ dephasing times. For two-qubit gates, engineering of the magnetic field gradient is possible through the composition of the qubits, resulting in different hyperfine strengths, while the tunnel coupling between the qubits and therefore the exchange interaction can be controlled via the separation. Aligning the qubits along the [110]-direction negates fluctuations in the strength of the exchange coupling as function of separation due to valley interference. The strength of the exchange coupling is high tuneable for asymmetric qubits (different electron number and / or number of phosphorus atoms).

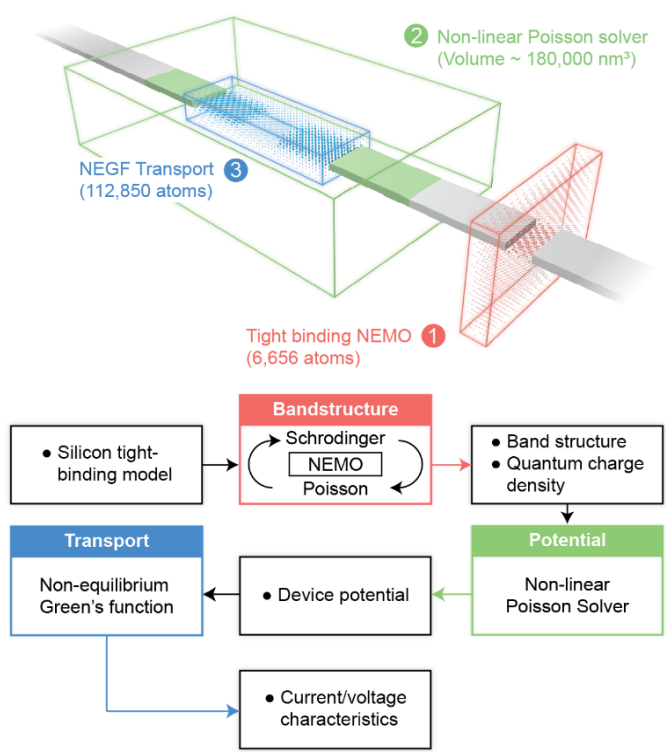

Figure 2: **Multi-scale modeling in atomically precise quantum devices.** This case illustrates the different domain sizes used in modelling tunneling transport in a Si:P tunnel junction [DONNELLY2023]. The source and drain of the tunnel junction (solid grey) are modelled using a tight-binding method (red), the full 3D potential of the channel is modelled using a non-linear Poisson solver (green), and the tunneling transport is modelled using the non-equilibrium Green's function (NEGF) formalism (blue). In each case, we show the number of atoms or volume considered in each modelling technique.

## Section 3.2 – Quantum control of donor spins

Holly G. Stemp and Andrea Morello

School of Electrical Engineering & Telecommunications, UNSW Sydney, NSW2052, Australia

**Status**

The idea of encoding quantum information in the nuclear spin $I = 1/2$ of phosphorus ($^{31}$P) donors in silicon [1] constituted a genuine watershed in practical quantum information science. It afforded the prospect of building atom-like physical qubits within a platform compatible with industry-standard semiconductor manufacturing. The simplicity of the donor spin Hamiltonian and the expectation of long spin coherence times [2] encouraged the pursuit of this goal.

Early results on single-donor devices were obtained in ultra-scaled fin-FET transistors [3]. Quantum control and readout of donor spin qubits became possible through an architecture that integrates the donors with a single-electron transistor for charge readout and spin-dependent tunnelling, and an on-chip microwave antenna to drive electron (ESR) and nuclear (NMR) spin transitions (Fig. 1a) [4]. This architecture has been the workhorse of both ion-implanted [5] (Fig. 1b) and has been subsequently adopted in STM-fabricated donor devices [6]. The first single-shot spin readout in silicon [7], electron spin qubit [5] and nuclear spin qubit [8] were demonstrated using this scheme.

The adoption of isotopically enriched, spin-zero $^{28}$Si substrates for ion-implanted devices enabled record coherence times (up to 35.6 seconds [9]), high-fidelity (close to 99.99% [10]) single-qubit gates, and strong violation of Bell's inequality between electron and nuclear spin in a $^{31}$P donor [11]. More recently, a pair of ion-implanted $^{31}$P donor nuclei was used to demonstrate 1- and 2-qubit logic operations with fidelity >99% [12].

Donor qubits nowadays offer multiple physical mechanisms for coherent quantum control. The natural one is magnetic resonance (NMR and ESR), as used in all early demonstrations in ion-implanted donors [5], [8], [9] and STM donor clusters [13]. Shallow donors, however, also possess a sizable electric polarizability, which enables electrical control ('Stark shift') of the electron-nuclear hyperfine interaction. Stark shift control with local gates was used for coherent spin control in an always-on, global oscillating magnetic field [14]. Driving the Stark shift at the resonance frequency corresponding to the electron-nuclear 'flip-flop' transition creates a new type of single-atom spin qubit that can be controlled via electric dipole spin resonance (EDSR) [15]. The large electric dipole associated with this qubit may be used for coupling to microwave resonators [16].

Using ion implantation, all Group-V donors can be introduced into silicon. The $^{123}$Sb donor possesses a quadrupolar nucleus with $I = 7/2$. A local oscillating electric field can drive nuclear electric resonance (NER) transitions by modulating the nuclear quadrupole coupling [17]. NMR, ESR, NER and EDSR driving mechanisms can be combined to give magnetic and electric access to the whole Hilbert space of the donor [18] (Fig. 2).

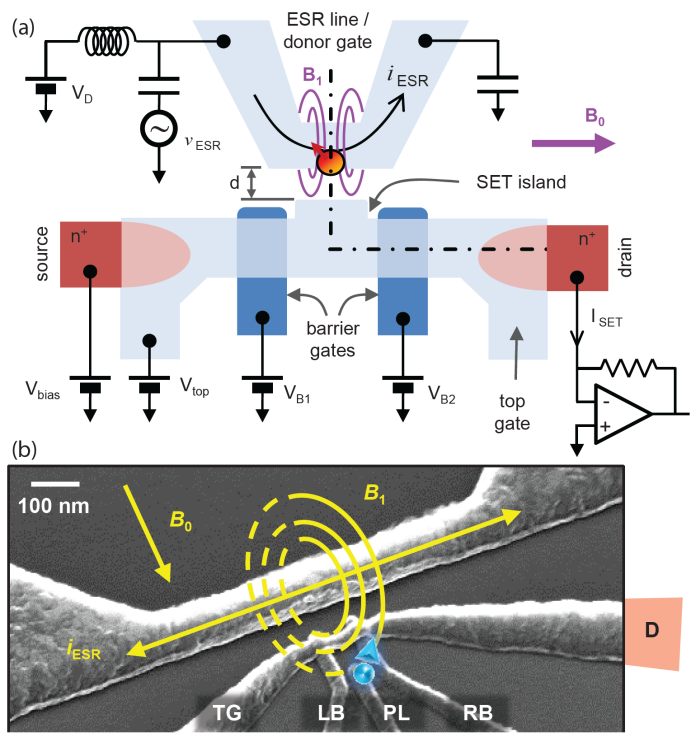

**Figure 1** (a) donor device control and readout architecture, integrating a microwave antenna and a single-electron transistor for qubit readout [4]. (b) Ion-implanted single-donor device emobyding the architecture [5].

**Current and Future Challenges**

*Growing the Hilbert space*: As with all technologies aimed at the development of quantum computing hardware, the key challenge with donor spins is growing the size of the computational (Hilbert) space to execute useful quantum algorithms at scale. This requires building large arrays of coherent, addressable single atoms, and engineering their interaction. Donor atoms have the benefit of being nominally identical quantum objects, thus less prone to the local variability in physical parameters (orbital and valley splitting, *g*-factors, etc) that can affect lithographic quantum dots. Small variabilities exist, due to local changes in electric field and strain, and these can be exploited to distinguish otherwise identical atoms, placed in a global driving field [14]. The key challenge is engineering inter-donor coupling. The tightly confined electron wave function, with a Bohr radius of order 2 nm, and the valley oscillations caused by the silicon band structure [19], make the exchange interaction strongly dependent on the precise donor position. This observation calls for creative methods to engineer interactions, which do not depend on the precise value of the exchange [20].

*Noise and quantum performance*: At the single-donor level, coherence times and gate fidelities are exceptionally high when the atoms are placed in isotopically enriched $^{28}$Si. All published data to date were obtained in materials with residual 800 ppm $^{29}$Si concentration, leaving room for further improvement. However, multi-qubit coupling inevitably requires involving the charge degree of freedom, which is more susceptible to noise. Recent data on weakly ($\approx$ 10 MHz) exchange-coupled ion-implanted donors shows no effect of exchange on electron spin coherence [20], which is very promising. The strong exchange limit does reveal the effect of charge noise [21], but the use of all-epitaxial, crystalline environments can mitigate the problem [22].

The electrically-driven donor systems, i.e. electron-nuclear flip-flop qubits and quadrupolar nuclei, are also sensitive to electric noise, due to the induced electric dipole and the nuclear quadrupole moment, respectively. However, the reduction in coherence time compared to single electrons or $I = 1/2$ nuclei is less than an order of magnitude [15], [18].

**Advances in Science and Technology to Meet Challenges**

*Growing the Hilbert space*: a promising direction in efficiently growing the Hilbert space of donor spin systems is the use of intrinsically high-dimensional atoms, such as $^{123}$Sb. The combined electron-nuclear Hilbert space has 16 dimensions: 2 from the $S = 1/2$ electron, 8 from the $I = 7/2$ nucleus (Fig. 2) [18]. The quantum control of such enlarged Hilbert space has become routine thanks to the latest progress in FPGA signal generators, capable of synthesising multiple frequencies while keeping a well-defined phase relation between them. Access to high-spin nuclei opens the possibility of encoding error-corrected logical qubits [23], which exploit the size and symmetry of the Hilbert space, without requiring engineering interactions between multiple and separate physical qubits. Multi-qubit operations between encoded logical qubits will pose similar demands to gates between simple qubits – i.e. control of exchange or electric dipole interactions – but require more complex multi-frequency pulses. Such pulses have been recently used to create and manipulate Schrödinger cat states of the $^{123}$Sb nucleus [24]

*Noise and quantum performance*: Ongoing efforts are underway to provide the community with high-enrichment $^{28}$Si substrates, with residual $^{29}$Si concentrations below 3 ppm [25]. Integrating these materials with the qubit control structures will enable exceptional coherence times and control fidelities at the single-qubit level. Moving forward, progress in charge noise, interface quality and robust control pulses will be necessary to extend the performance enhancements to multi-qubit operations. High-spin donor nuclei may be used as accurate detectors of device noise, since it is possible to rigorously separate out the magnetic from the electric component of the noise [18]. They also provide atomic-scale information on lattice strain [17], which is a key parameter in all semiconductor quantum (and classical) devices.

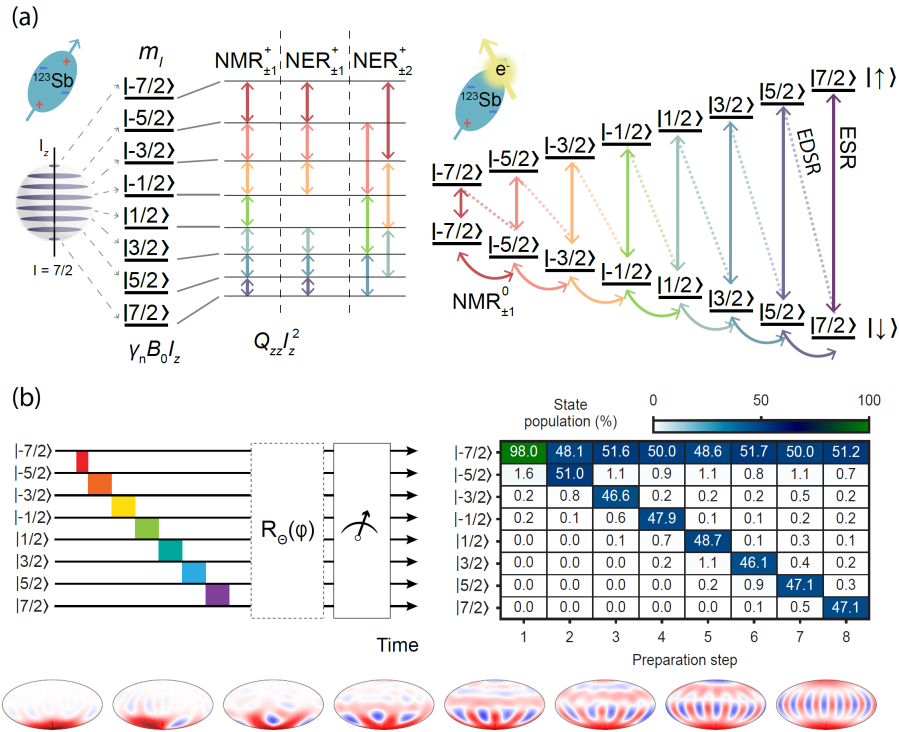

**Figure 2** (a) Energy level diagram of the 123Sb donor atom, comprising a 16-dimensional Hilbert space that can be controlled by magnetic (NMR, ESR) and electric (NER, EDMR) fields [18]. (b) Creation of a Schrödinger cat state of the 123Sb nucleus, by applying a sequence of coherent multi-frequency pulses. The state populations, and the corresponding spin Wigner functions, are shown for every step of the sequence [24].

**Concluding Remarks**
Despite being a rather 'niche' choice of physical quantum hardware compared to e.g. lithographic quantum dots, donor atoms remain at the forefront of the quest for performance and scalability. For the ion-implanted ones in particular, much of the progress in device fabrication, material and interface quality pursued in the wider semiconductor qubit field will transfer directly to the donor devices. On the flip side, the exceptional quantum coherence and well-understood spin Hamiltonian of the donors may become a precious tool to support the development of the broader semiconductor quantum hardware. The prospect of encoding error-corrected logical qubits in high-spin donors provides an exciting shortcut to the realisation of useful quantum hardware in the medium term.

**Acknowledgements**
The authors acknowledge support from the Australian Research (grants no. CE170100012, DP210103769) and the US Army Research Office (contract no. W911NF-23-1-0113). The views and conclusions contained in this document are those of the authors and should not be interpreted as representing the official policies, either expressed or implied, of the Army Research Office or the U.S. Government. The U.S. Government is authorized to reproduce and distribute reprints for Government purposes notwithstanding any copyright notation herein.

## Section 3.3 - In-situ quantum measurements

Benoit Voisin[1,2] and Sven Rogge[2]

[1] Silicon Quantum Computing Pty Ltd, UNSW Sydney, Kensington, New South Wales, Australia
[2] Centre of Excellence for Quantum Computation and Communication Technology, School of Physics, UNSW Sydney, Kensington, New South Wales, Australia

**Status**
Atomic devices see their electronic properties and performance relying on a few atoms only. While these devices represent a great prospect for quantum applications, this ultimate scaling also comes with enhanced sensitivity to the local environment such as precise atom positioning, strain or interface disorder. These effects must be controlled to unlock scalable architectures, however they cannot be easily captured in usual transport spectroscopy experiments as correlating measurements with the precise atomic-scale details of the device can be challenging. Instead, microscopy techniques can provide local information on single atomic devices [1-4], to become a useful resource to optimize their geometry and performance.

We focus here on scanning tunnelling microscopy, as it provides single atom resolution and the ability to directly probe the wave function of quantum states embedded in a crystal host. This technique has been used to image defects and dopants at semiconductor surfaces, first in III-V materials and more recently in group IV. More precisely, the wave function of single As and P dopants embedded below the silicon surface were imaged, revealing a direct measurement of the mass anisotropy and valley interference in silicon [5]. This interference pattern can be used to determine the position of the dopant with exact lattice site precision [6], and observing the valley interference between a pair of dopants revealed a preferential crystallographic direction to mitigate their impact on the exchange interaction [7].

While STM provides atomic resolution, it has usually been limited to simple vertical geometries with conductive substrates. This situation contrasts with quantum devices which benefit from insulating substrates and local gating for coherent manipulation. A stepping stone towards closing this gap was recently taken with the ability to simultaneously image and gate the wave function of engineered dopant-based quantum dots in undoped silicon using a light-assisted protocol at 4K [4] (see **Fig.1**). We discuss below the next steps and challenges ahead to extend this in-situ atomic device imaging framework. This notably includes probing the spin degree of freedom at sub-Kelvin temperatures and perform coherent manipulation, and scaling-up towards correlated 2D arrays, both fabricated and measured with an STM in-situ.

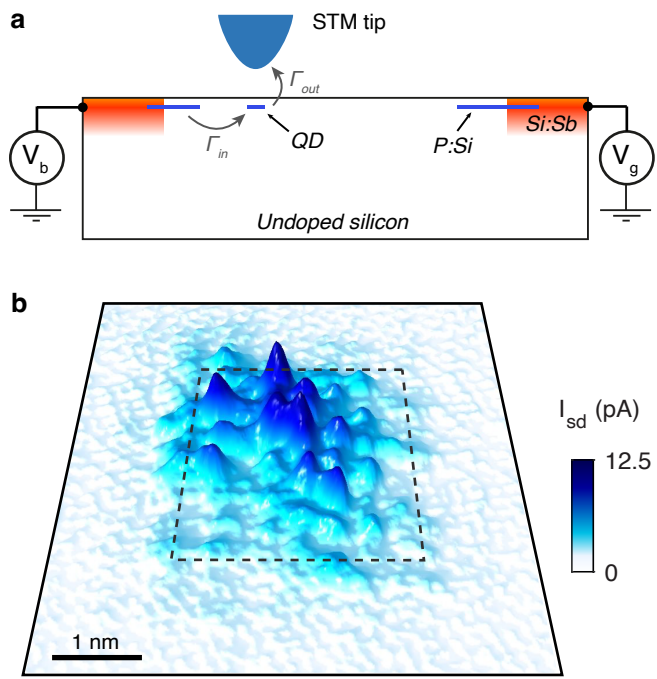

**Figure 1 In-situ quantum state imaging of an atomic-scale quantum device in silicon. a,** Electronic transport occurs between an in-plane source reservoir, made of a Sb-doped lead extended with a phosphorous lead designed with atomic precision using STM lithography, a quantum dot (QD) also defined with atomic precision, and the STM tip. Another in-plane electrode acts as a gate to control the electrochemical potential of the QD. **b,** Quantum state image of an atomically precise QD. The 3D image is a representation of the electronic current $I_{sd}$ flowing from the source electrode to the STM tip, which can be associated with the charge distribution of the QD at the silicon surface convoluted with the tip orbital. The charge distribution is well included within the QD area designed with STM lithography (dashed square).

**Current and Future Challenges**

Single atom and molecule resolution has been achieved using AFM and SGM [8]. In parallel, they also have demonstrated operations on quantum devices, so the next goal is to combine these realisations in the context of an operating atomic-scale device.

The main direction for STM is to develop the ability to probe the spin degree of in atomic-scale quantum devices, leveraging the achievements using magnetic atoms. In this framework, the spin was recently coherently manipulated in coupled atom systems prepared with atomic precision [9], however, the coherence of these systems has so far been limited due to the strong interactions between the quantum state and the substrate or the STM tip. Realising this in-situ quantum manipulation technique in a semiconductor, where coherence can be maximised using the local gating ability recently developed, will require to devise appropriate read-out protocols and an electrically driven spin resonance mechanism either through a spin-orbit qubit system or a magnetically polarised tip. An important challenge will be to characterise the tunnel and exchange coupling interactions in this experimental situation.

The second main direction is to develop and probe large-scale arrays. Dopants in silicon are a strongly correlated system, with charging energies that can be tuned from a few meV to above 100meV depending on the size of the dopant-based quantum dot. Tunnel couplings $t$ can easily exceed 1meV using STM atomic precision dopant placement, representing a unique opportunity to achieve the quantum regime $T \ll t$ for temperatures $T$ below 1K [10]. In a quantum simulation context, the STM can be used to fabricate a large 1D or 2D array of dopant-based quantum dots and to probe its ground state by scanning the quantum state of each site. The challenge is to define clear measurement protocols to probe correlations between the different sites to be able to demonstrate the existence of many-body correlated states in the array [11]. An effort will also focus on mitigating short and long-range disorder inherent to semiconductors.

Finally, another direction of research will focus on implementing a hybrid integration between the in-plane doped gate geometry and other types of dopants or defects. Of particular interest are boron atoms [12], which present a strong spin-orbit coupling amenable to electrically-driven spin manipulation, and optically-addressable erbium atoms [12]. In the latter case, the in-situ quantum imaging could provide direct information on the complex site symmetry physics existing in this system.

**Advances in Science and Technology to Meet Challenges**
A major recent advancement in microscopy technique has been the ability to functionalize the tip [9], enabling to improve the tool spatial resolution and to probe the spin degree of freedom of magnetic atoms. Its extension to semiconductor systems is therefore a necessary step, however it is challenging given the very stringent cleanliness requirement for in-situ semiconductor fabrication with a high sensitivity to magnetic or molecular contamination. Also, the very low temperatures, below 100mK, achieved in-situ for magnetic systems were made possible thanks to the high thermal conductivity associated to these electrically conductive substrates. Achieving electronic temperatures below 100mK for semiconducting and insulating substrates while preserving STM mechanical stability will be technically challenging. Bridging the gap between quantum devices measured in conventional cryogenic systems and in-situ will also require installing adequate low-temperature high-frequency components for high-fidelity coherent manipulation. This is not trivial given the compactness of UHV ultra-low temperature STM systems. Another area of improvement will be to enable multi-terminal STM sample holders for in-situ quantum devices, with enhanced risks of contamination and failure during annealing steps that will have to be carefully mitigated (see Figure 2).

Finally, the in-situ quantum imaging framework will benefit from the recent development of novel single atom quantum devices based on spin-orbit quantum systems or erbium atoms. Similarly to what has been achieved for phosphorus dopants, the challenge here will be to modify the ion implantation or incorporation techniques for the atoms to remain close to, or on the silicon surface, while maintaining a clean and flat enough surface amenable to quantum state imaging.

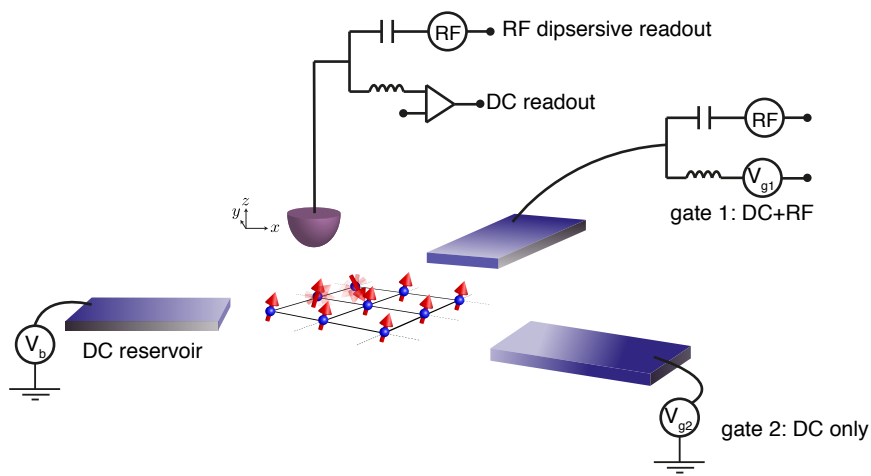

**Figure 2 In-situ imaging: long-term prospects.** A two-dimensional array of dopant-based quantum dots is fabricated with atomic precision using STM lithography. A set of controlled in-plane electrodes is used to tune the filling number of the array. The STM tip, which can be functionalised, is used to probe the ground state and correlations within the array. Both the STM tip and electrodes can also be used to mediate coherent spin manipulation and read-out. The in-situ device should operate in finite a magnetic fields $B$ to probe the spin degree of freedom and associated low temperatures ($T$<100mK) to achieve $k_B T \ll g\mu_B B$.

**Concluding Remarks**

Benefiting from its natural atomic spatial resolution, scanning tunnelling microscopy has been pivotal to unveil the complex valley physics associated to single and coupled donors in silicon, providing crucial information to optimise atomic-scale quantum devices. This work was extended to a full device configuration, where the dopants quantum states can be simultaneously imaged and controlled using side gate voltages directly in-situ. This represents a stepping stone towards probing the spin degree of freedom and achieve highly coherent in-situ manipulation in semiconductors. Towards this objective, future work will focus on the required technical developments to achieve low temperature and finite magnetic fields. Benefiting from the STM atomic precision fabrication and the versatility of the STM tip as scanning probe, this platform can also naturally be extended to large 2D array of strongly interacting dopants, to form a unique platform to investigate artificial correlated matter in the quantum regime. Finally, this in-situ quantum imaging framework could be implemented with novel sites of interest for atomic-scale quantum devices such as colour centres to explore the rich physics of hybrid light-matter systems in semiconductors.

**Acknowledgements**

We acknowledge support from the ARC Centre of Excellence for Quantum Computation and Communication Technology (CE170100012), an ARC Discovery Project (DP180102620) and Silicon Quantum Computing Pty Ltd.

# Section 3.4 – Silicon Dangling Bond Quantum Dot Devices

Robert A. Wolkow[1], Lucian Livadaru[1] and Jason Pitters[2]

[1]University of Alberta, Department of Physics, Edmonton, Canada T6G 2E1
[2]National Research Council of Canada, Quantum and Nanotechnologies Research Centre, 11421 Saskatchewan Drive NW, Edmonton Canada T6G 2M9.

**Status**

This article is about the extraordinary prospects for realizing spin and charge devices based upon Dangling Bonds, DBs. Elsewhere in this collection DBs are described as anchor points for spin possessing atoms such as phosphorus or arsenic. But advances in understanding of DB electronic properties, together with the new ability to fabricate error-free ensembles consisting of >100 DBs have revealed the DB to be a very attractive building block of quantum devices and also quantum enabled ultra-low power and ultra-fast classical circuitry[1].

The Si DB is a gap state occupied by 0, 1 or 2 electrons. Paired DBs are strongly tunnel coupled. Symmetric-antisymmetric splitting is of order 100 meV, exceeding the tunnel interaction among conventional Si quantum dots by ~1000x. Binary Atomic Silicon Logic, BASiL, has been demonstrated[1]. A Google TPU based on BASiL rather than CMOS is predicted to reduce energy consumption 10,000 fold at similar speeds[2]. While fabrication rates must increase by a factor 100,000 for manufacturing of complex devices to become feasible, very substantial improvements (~1000 fold) have been demonstrated and the first commercial tool is on the horizon[3] . Far simpler quantum devices of extreme small size, weight and power, that are operable at ambient conditions are expected to emerge very shortly, for example a quantum random number generator, QRNG[4].

The singly occupied DB state is paramagnetic, nominally like a P atom in Si. Compared to a dopant atom, the paramagnetic DB is easily prepared, unerringly placed, simply interfaced to atomic wires and does not require epitaxial overgrowth. DBs do however require vacuum encapsulation. Whereas, spin characterization of DBs is in its infancy, DB charge qubits have been proposed and modelled. Inspired by[5], highly noise immune quadrupolar DB qubits are currently under study. An extensive review has just been published[6].

A proof of principle of an atom-defined single electron transistor, SET, has been shown. The trapped charge effects that limit other types of SETs are eliminated. An Inherently large Coulomb blockade energy allows for room temperature application. Variants of that configuration will be juxtaposed with BASiL logic elements and the QRNG to transduce signals from atom-scale to CMOS.

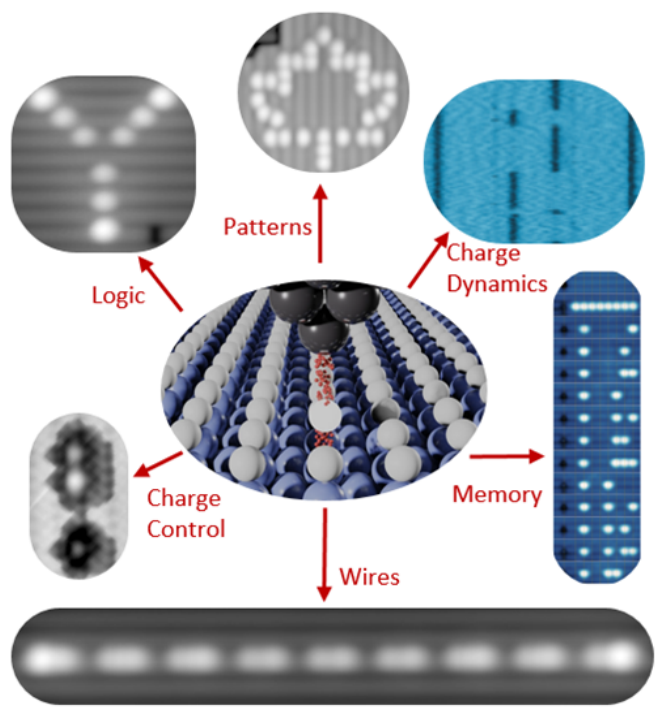

**Figure 1** Scanning Tunneling Microscopy hydrogen desorption lithography can create DB patterns for various functions.  Used with permission of Springer Nature BV. from [1] permission conveyed through Copyright Clearance Center, Inc.  Used with permission of Springer Nature BV from [7], Copyright 2018; permission conveyed through Copyright Clearance Center, Inc.  Reprinted in part with permission from [4]. Copyright 2018 American Physical Society. https://doi.org/10.1103/PhysRevLett.121.166801.  Reprinted with permission from [8]. Copyright 2009 American Physical Society. https://doi.org/10.1103/PhysRevLett.102.046805

**Current and Future Challenges**
Exactingly inserted atoms are required to create reproducible, variance free, dopant-based quantum devices. Imperfections result from; 1) errors in H lithography, 2) the stochastic process of dopant precursor attachment, 3) diffusion, and 4) non crystallinity created during Si overgrowth. These challenges provide motivation to explore simpler, more widely deployable DB-based devices as an alternative to dopant-based ones. Today, inserted dopant patterning is limited to ~1 nm uncertainty. Only DB patterning can presently be rendered perfectly with true atomic precision.

Successful interfacing of atoms to macroscopic electrodes has predominantly involved saturation-doped lines of P atoms to create lines that are metallic at cryogenic conditions. To be activated the P atoms must be overgrown with silicon and sufficiently annealed to achieve tetrahedral crystalline lattice positions. For some planned devices, more exactly defined and more spatially abrupt wire/ device interfaces are required, creating a need for a different approach to wiring-up atomic devices.

Preparation of defect free silicon surfaces remains a challenge. Standard lithography steps, such as those required to implant patterned dopants exacerbates the surface cleanliness problem. Improved procedures are required.

 DBs have been extensively studied and shown to be richly capable but must be maintained in ultra-high vacuum. Robust encapsulation is required to further explore and ultimately to deploy DB-based devices.

Error free DB printing is now in hand. However, speed of DB printing is limited by inaccuracy of scanned probe placement. Far more rapid, absolutely accurate probe positioning is required.

Methods of preparation, measurement and controlled coupling of DB spins is required to create spin qubits and related quantum sensors. Commensurately small-scale spin filter and SET readout devices are also required.

**Advances in Science and Technology to Meet Challenges**
DB wires will be deployed where more sharply spatially, and energetically defined properties are required than can be delivered by dopant wires. DB wires can provide exactly placed, as-designed coupling to other atom-defined entities such as quantum dots. DB wires are themselves quantum entities with exactly reproducible discrete quantum modes. Whereas nanoscopic dopant wires contain a scattered distribution of dopant atom placements and density and therefore exhibit "noisy band"-like conduction states tailing toward bulk band edges, DB wires crucially have conducting states deep within the bulk silicon band gap, thereby reducing leakage via substrate conduction. Leakage creates cross talk, heating and decoherence effects in devices. Designed local dopant concentration and electrostatic gating can achieve connection to DB wires as desired. The DB building block also enables a different kind of wire, a binary wire that cannot be made of placed dopants. The transmission of binary state, without the use of conventional current, effectively eliminates Joule heating.

Tests of encapsulation through wafer bonding are underway. It is expected that permanent exclusion of air and other gaseous contaminants will be routinely achieved, creating deployable atom-defined devices of many year lifetimes.

Atom scale printing must become perfectly error-free and fast. How fast? At least, fast enough to print ~100s atom devices on a minute's time scale. At best, fast enough to print billions of atoms in 10 hours. Four areas for improvement:
1) Defects in tip structure must be quickly and automatically detected and repaired without human intervention[9]. Much progress has been made in atom defined tip fabrication[10].
2) Rapid tip positioning with $10^{-11}$ m uncertainty is required. A 3-dimensional UHV compatible absolute positioning device has been developed[3].
3) Rapid atomic editing by controlled H replacement is required to enable perfect patterning [7].
4) Silicon substrates of reduced defect density are required. Revised sample cleaning procedures can achieve this.
5) Automated mapping and categorising of pre-existing defects is needed, as is defect avoiding circuit layout. Circuits and constituent components must allow for a degree of morphing of detailed layout[11, 12].

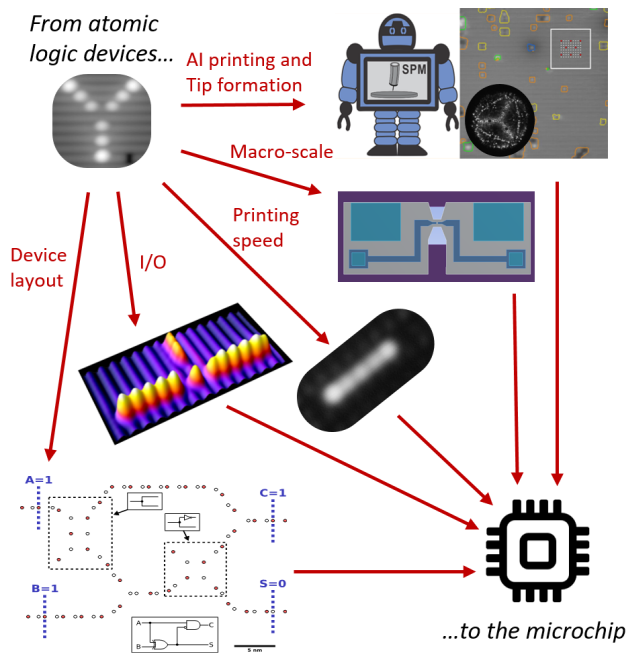

**Figure 2** Future developments to transform devices to functioning circuits. Used with permission of Springer Nature BV. from [1] permission conveyed through Copyright Clearance Center, Inc. Reprinted with permission from [9]. Copyright 2018 American Chemical Society. Reprinted with permission under a Creative Commons CC BY 4.0 License from [11]. Copyright 2020 IOP Publishing Ltd. https://creativecommons.org/licenses/by/4.0/. Reprinted from [10], 2013, with permission from Elsevier. Reprinted with permission from [13]. Copyright 2020 Stephanie Yong. Reprinted with permission under a Creative Commons CC BY 4.0 License from [12]. Copyright 2020 IEEE. https://creativecommons.org/licenses/by/4.0/

**Concluding Remarks**
With respect to silicon dangling bonds, the long-standing goal of Atomically Precise Manufacturing, APM, is at hand. Refinements will be made but the essential challenges have been met. Truly identical solid-state entities can be made, enhancing scalability by eliminating calibration and tuning. Unerringly defined interconnects can also be made of DBs. DB structures can embody quantum and classical ultra-low power and ultra-fast devices. DB control circuitry can also enhance control over dopant spins. Printing with DBs is slow in that it is serial but fast because all components are completed in one pass. A truly 2-dimensional circuitry results. Multilayers and vias are eliminated. AI-based automation makes probe lithography a practical path. A new error free scanner will greatly speed up atom writing. Error free printing of many thousands of atoms is just around the corner. Parallelization and other fundamentally faster writing processes are foreseen.

**Acknowledgements**
Funding has been supplied by the National Research Council of Canada, Alberta Innovates Technology Futures, Natural Sciences and Engineering Research Council of Canada and Compute Canada.

# Section 4.1 – Deterministic dopant incorporation

Taylor J. Z. Stock,[1,2] Steven R. Schofield[1,3] and Neil J. Curson[1,2]

[1] London Centre for Nanotechnology, University College London, 17-19 Gordon St, WC1H 0AH, UK
[2] Department of Electronic and Electrical Engineering, University College London, WC1E 7JE, UK
[3] Department of Physics and Astronomy, University College London, WC1E 6BT, UK

**Status**

Motivated by the materials requirement of the Kane quantum computer architecture, the incorporation of individual phosphorus dopant atoms in the silicon (001) surface with near-atomic precision was first demonstrated in 2003 [1]. A single phosphorus atom was positioned with 1 nm precision and a pair of phosphorus dopants were positioned along a single dimer row separated by approximately 12 nm. This proof-of-principle demonstration was the first step in enabling the fabrication of a wide variety of single donor devices, yet none so far have involved large numbers of deterministically positioned single donors. Key to such large-scale deterministic donor incorporation is the requirement that individual dopants can be incorporated at lithographically defined adsorption sites with near-perfect reliability. However, the thermodynamics of phosphine ($PH_3$) on Si(001) appears to impose severe limits on this incorporation, with a demonstrated probability of only 63 ± 10% [2], a serious limitation for device scale up.

While there may be potential to exceed the thermodynamic limit for phosphorus incorporation in silicon through, for example, the use of STM tip induced incorporation (Section 4.2), a promising alternative approach is to change the material system from phosphine on Si(001) to arsine ($AsH_3$) on Si(001) [3, 4] or even arsine on Ge(001) [5]. Unlike phosphine, arsine has a rapid dissociation pathway upon adsorption to the silicon surface and is robust against recombinative molecular desorption during the incorporation anneal. It has been shown that arsine incorporation at a lithographically defined four silicon lattice adsorption site occurs with a 97 ± 2% probability. Using this technique, the first defect-free (2x2) array of individual dopant atoms in silicon was fabricated [3]. It is expected that using an iterative lithography/dose cycle, a single atom incorporation probability of up to 100% could be achieved.

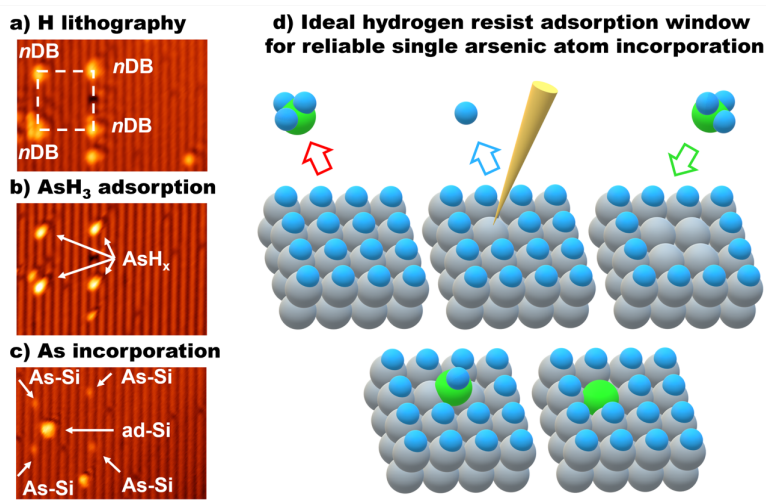

**Figure 1: Single atom control of arsenic incorporation in silicon**: Panels a–c) show STM images from each step of fabricating a 2 × 2 single-atom arsenic array in silicon. The illustrated process steps are: a) STM lithographic patterning of the hydrogen resist; b) room temperature $AsH_3$ adsorption; c) thermal anneal for activation of substitutional arsenic incorporation. Labelled features are: $n$DB = lithographic dangling bond patch, ad-Si = silicon adatom, As-Si = silicon-arsenic heterodimer, and $AsH_x$ = arsine molecule fragment ($x$ = 0 to 2). Panel d) illustrates these fabrication steps schematically, beginning with the unpatterned resist, and presents the ideal 4 atom silicon-site adsorption window that can be used to ensure incorporation of single arsenic atoms with 97±2% probability. Imaging parameters: a) −2 V, 40 pA; b,c) −2 V, 100 pA. Lithography parameters: 3.5 V, 3000 pA, 50 nm·s$^{-1}$. $AsH_3$ dosing: 5 × 10$^{-9}$ mbar × 10 min. Incorporation anneal: 350 °C × 1 min. Panels a)-c) reproduced from Adv. Mater.2024, 2312282 under the CC BY license (https://creativecommons.org/licenses/by/4.0/).

The electrical characteristics of two-dimensional arsenic δ-layers have been shown to approach those of similarly prepared phosphorus δ-layers [4]. Soft x-ray angle resolved photoemission measurements have shown that arsenic δ-layers exhibit better thickness confinement than identically-prepared phosphorus ones [6]. On the germanium (001) surface, arsenic atoms have been shown to incorporate spontaneously at room temperature after exposure to arsine [5]. This exciting result removes the requirement for an incorporation anneal entirely and work is ongoing to demonstrate compatibility with lithographic confinement.

**Current and Future Challenges**
Unlike P/Si(001) where devices are now routinely fabricated and measured, for all other material combinations, including As/Si(001), the immediate challenge is to demonstrate functioning electrical devices incorporating single dopant atoms and arrays.

The electrical transport properties of small arrays of individual arsenic dopants must be studied, with the data expected to reveal the presence of Hubbard bands and the Mott gap [7]. Increasing the size of the arrays and varying the array motif will open the opportunity for the simulation of various quantum systems and here in particular the input from theoreticians (see Section 2.1) will be especially important for the design and interpretation of future experiments.

The materials pallette for atomic-scale devices must also be further expanded. In addition to phosphorus and arsenic, good progress has been made with the acceptor species, boron. The fabrication of nanoscale bipolar devices [8] and boron delta-layer nanowire devices [9] have been achieved. Experiments with boron are still far off the single dopant limit, however, theory suggests it can be reached [10]. Dopants with nuclear spins of $I \geq 1$, such as arsenic (nuclear spin $I = 3/2$), antimony (nuclear spin $I = 7/2$) and bismuth ($I = 9/2$), and associated nuclear quadrupole moments, are exciting for the additional degrees of freedom they provide for performing quantum logic operations, and coherent electrical control of a single ion-implanted antimony nucleus in silicon has been demonstrated. Work has begun aiming to isolate individual bismuth dopants on silicon with bismuth trichloride showing promise [11]. As yet, no work has been done to develop hydrogen lithography compatible precursors for optically active species like erbium.

Critical to the successes of all the precursor development described above and any future work is a detailed understanding of the surface chemical and thermodynamic processes. Only through the combination of atomic resolution STM images and detailed density functional theory (DFT) calculations can such information be obtained. DFT provides not only optimised geometries, formation energies, and simulated STM images for direct comparison to experiment, but also transition state energetics for a full description of chemical reaction pathways that often occur on timescales much faster than can be accessed by STM.

**Advances in Science and Technology to Meet Challenges**
The use of arsine in place of phosphine is motived by the necessity of creating defect free large-scale donor qubit arrays. In addition to this change in material system away from phosphine/silicon, several other technological advances will also be required to implement large-scale deterministic dopant incorporation.

1) Resist quality. To avoid the unintentional creation of impurity defects or spurious dopant incorporation at unintended sites, we must address the quality of the resist layer. Process improvement for the quality of the hydrogen (or halogen) termination may be possible. Alternatively, strategic use of multiple precursor gases might be exploited to selectively enable or block adsorption as discussed in Ref. [3].

2) Precisely controlling, and verifying, the dopant positions in three-dimensions. Minimising dopant segregation during overgrowth and annealing will be crucial, and it cannot be assumed that segregation will mirror that of dense 2D delta-layers, which are well studied [6], as the solid solubility of dopants in the different density regimes is not comparable. Additional development of locking layer and rapid thermal annealing protocols will be required. The diagnostic technique with the most promise for directly imaging arrays of individual dopants is atom probe tomography (APT), which has, for example, been used to image multiple layers of phosphorus dopants in germanium [12,13].

3) To target and manipulate the states of individual elements of dopant arrays, global and local gating (either back gates or top gates) are required to control the chemical potential in which the dopants reside, and on-chip antennas are needed to manipulate individual qubit spin states. The orbital states of donors can be excited using terahertz radiation, with germanium a particularly suitable host as the excitation energies are accessible by lab-based quantum cascade lasers. As the physical extent of the excited orbital states extend over many more lattice sites than the associated ground states, terahertz offers the possibility of optically switching on and off dopant interactions of an array.

**Concluding Remarks**
The Kane quantum computer proposal in 1998, to create qubit arrays of individual phosphorus donors in silicon, provided the ideal application for STM-based atomic manipulation in semiconductors. The burgeoning field of hydrogen depassivation lithography and its use to control adsorption to the Si(001) surface was quickly adopted and individual phosphorus atoms were positioned in silicon in 2003. The first single atom transistor was fabricated and measured in 2012 and numerous other exciting and foundational advances and devices have been made since then as described elsewhere in this article. Yet the goal of deterministic large-scale single-atom arrays has not been achieved. We argue here that progress in this direction may require the move away from phosphine/Si(001) to other material combinations. Arsine/Si(001) in particular has shown remarkable properties that may lead to a breakthrough in scalability of donor qubit arrays. In parallel, much work remains for the development of other material combinations including acceptor and optically-active species, and alternative hosts like germanium.

**Acknowledgements**
We thank G. Aeppli, D. R. Bowler, P. C. Constantinou, E. Crane, Y. Ekinci, A. J. Fisher, S. Fearn, E. V. S. Hofmann, D. Kazazis, A. Kölker, J. Li, D. R. McKenzie, M. Muntwiler, V. N. Strocov, C. A. F. Vaz, and O. Warschkow for their roles in the work described here. The project was financially supported by the Engineering and Physical Sciences Research Council (EPSRC) projects EP/M009564/1, EP/R034540/1, EP/V027700/1, and EP/W000520/1, Innovate UK project UKRI/75574, the EPSRC Centre for Doctoral Training in Advanced Characterisation of Materials (EP/L015277/1), and the Paul Scherrer Institute (PSI).

## Section 4.2 – Tip-induced deterministic P incorporation

Jonathan Wyrick, Pradeep Namboodiri, Fan Fei and Richard Silver

Atom Scale Device Group, National Institute of Standards and Technology, Gaithersburg, MD 20899, USA

**Status**

Early work from Fueschle [1] showed that scale-down of the standard STM-based lithography process to adsorb $PH_3$ into lithographic patches consisting of 3 consecutive Si dimers on a single dimer row (figure 1(a)) enables embedding of a single P atom with an incorporation yield of 70%, and was later corroborated by definitive study by Ivie et al. showing a yield of 63% [2]. The incomplete single-atom yield from the standard process meant that to achieve true single-atom precision fabrication, alternative methods would be necessary, one possibility for which is the use of STM tip-based manipulation of the adsorbed $PH_x$ precursors. This solution to the problem was motivated by work showing that H atoms could be controllably dissociated from $PH_x$ molecules adsorbed on a non-H-terminated Si (100) surface [3], and has now been demonstrated to be an effective method for achieving 100% single P atom incorporation yield [4].

At present, $PH_x$ precursor manipulation is achieved by adsorbing $PH_3$ into 1-dimer lithographic patches created using feedback-controlled lithography (FCL) [4], [5], yielding $PH_2$+H (figure 1(b)), followed by removal of H atoms that surround the adsorbed species (figure 1(c)). This enables feedback-controlled manipulation (FCM) of $PH_2$, using the same signals and control as FCL, to dissociate the 2 H atoms from the molecule, creating a lone P atom (figure 1(d)). Heating this final configuration to 350C results in 100% incorporation of the P atoms into the Si lattice, within +/- 1 lattice site from the intended location determined by the initial 1-dimer patch.

There are several important directions for further development of this method: adaptation to other molecular precursors to increase the number of atomic/molecular species that can be used to construct single-atom/molecule precision devices, manipulation to enable incorporation of more than one atom at a site, direct incorporation using the tip rather than heating, and, if possible, improvement of control to achieve exact site incorporation with no uncertainty. Making these advances will enable reliable fabrication of atomic-precision devices with greater complexity and variety of capabilities.

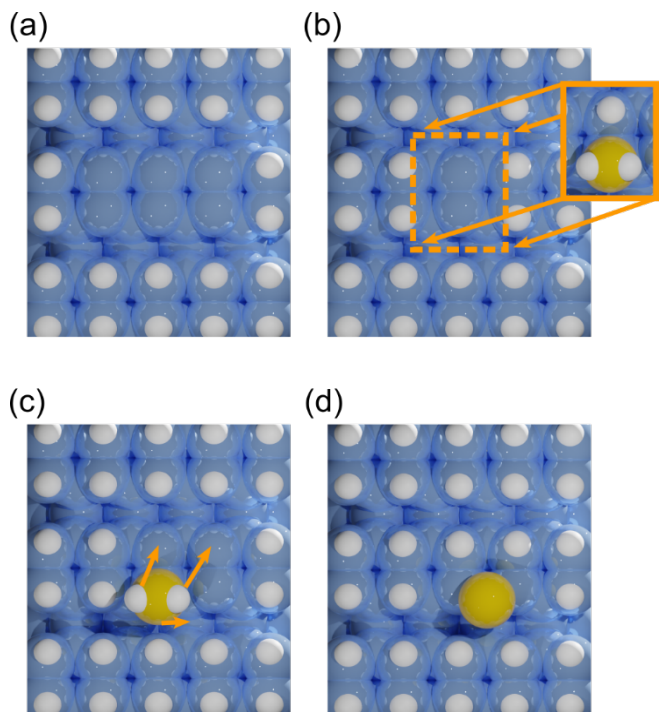

**Figure 1** Atomic configurations (white: H, blue: Si, yellow: P) during lithography and manipulation for single-atom incorporation. (a) 3-dimer lithographic patch ideal for single P atom incorporation when tip-based manipulation is not used, resulting in imperfect yields. (b - d) Process for tip-based manipulation, where (b) PH$_3$ adsorbs into a 1-dimer patch as PH$_2$+H (inset), FCL is used to remove 3 H atoms from the Si surface at the 3 top sites immediately above and to the right of the adsorbed PH$_2$ (c), then a final FCM step is performed causing atomic species to dissociate and move as shown by the arrows in (c), resulting in the lone P atom at a bridge site (d).

**Current and Future Challenges**

The key challenges of tip-induced manipulation and incorporation revolve around the fact that this technique is both time intensive, and human operator intensive. These challenges are inherited from hydrogen lithography and its application to device fabrication, but are amplified by the level of precision and near-perfection required for single-atom work. For the donor-dot device shown in figure 2, lithography, manipulation, and incorporation required more than 2 weeks of dedicated operator time.

A significant degree of expertise is necessary for the STM operator, as they must be able to identify and make informed decisions about defects and key features of the substrate, adsorbed molecules, and intermediates at each step of the fabrication process to perform appropriate tip-based manipulations. This further necessitates that the STM tip must be atomically perfect, both for the purpose of feature identification (the appearance of STM images drastically changes if the tip apex is not a single atom [6], [7]), and for FCL and FCM which rely heavily [4], [5] on the nature and geometry of the tip-induced potential [7] on the substrate. A great deal of effort is spent conditioning, and re-conditioning the STM tip during fabrication, particularly since the ability of the tip to perform manipulation necessitates that it directly interacts with the sample and is therefore equally likely to be modified itself during such procedures. As a result, errors can be introduced during device writing, and because of the need for near perfection, either the errors must be corrected by some means [8], or the operator must abandon the current effort, beginning device patterning anew elsewhere on the sample.

The likely direction for future challenges will be increasing the complexity of single-atom devices that can be fabricated. We can anticipate becoming more ambitious with the types of atoms and molecules we desire to incorporate, greatly increasing the domain of applications available. The manipulations required in some cases may go beyond the existing methods using tip applied bias/current, and may require multi-step reactions to complete, presenting new hurdles for this method to overcome.

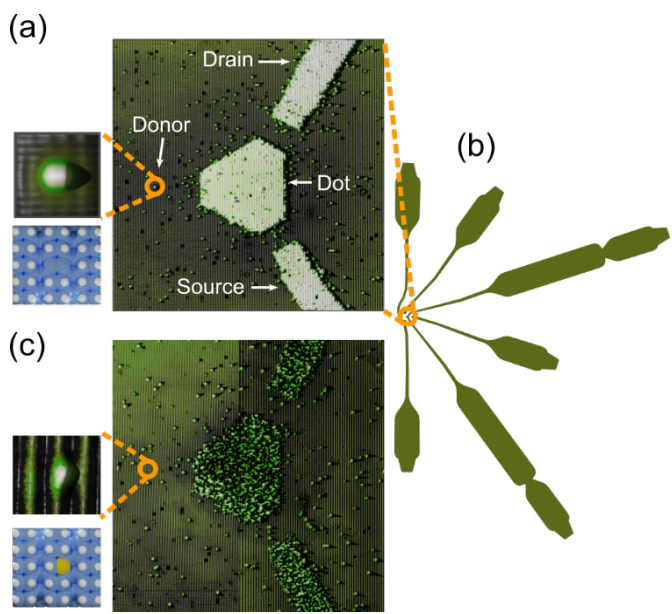

**Figure 2** Donor (single P atom) – dot (isogonal hexagon-island of P atoms) structure fabricated using hydrogen lithography and tip-based manipulation. (a) Donor-dot STM image taken at -2V sample bias immediately after lithography. The inset shows a zoom-in of the 1-dimer patch as imaged by the STM at +2V sample bias (top) and depicted as an atomic model (bottom). Bright protrusions seen away from patterned regions are defects, primarily single dangling bonds onto which $PH_3$ cannot adsorb. (b) Design of the full device including fan-out, used during lithography. (c) Donor-dot STM image (stitched together from 2 images) taken at -2V sample bias after incorporation, revealing ejected silicon islands directly over incorporated P in the dot and the source/drain leads, and a single incorporated phosphorus atom at the intended donor site (inset), -2V sample bias, with its representative atomic model (bottom). Length scales in STM images are given by the pitch of dimer rows of 0.77 nm.

**Advances in Science and Technology to Meet Challenges**
The fundamental problem of fabrication time and the need for operator expertise point strongly to the value of integrating automation into the tip-based manipulation process. Efforts to apply machine learning to problems such as feature detection [9] and tip conditioning [6] are already underway and show promising results, having been applied successfully to dangling bond device fabrication. These methods need to be applied to the broader problem of manipulating precursor molecules, tip-induced incorporation, and all other aspects of STM-based fabrication. Fully implementing machine learning and automation will accelerate the pace of device fabrication while minimizing the potential for human-introduced errors. Improvements to instrumentation that enhance STM tip quality and stability as it relates to patterning, imaging, and manipulation will also be key, as these will ease much of the burden currently taken on by the operator, or as advances are made, by the automated controller.

With regards to feature detection, an important component to tip-based manipulation has been simulated STM images enabling determination of precise atomic configurations of surface species during each step [4].  In expanding to a wider selection of precursors and manipulation techniques, this catalogue of identified configurations must also increase and must be integrated into automated processes.

Exploring alternatives to the currently used techniques for FCL and FCM will aid in overcoming the challenges presented both by new atomic/molecular species to be incorporated, as well as in efforts to achieve greater precision.  There is significant room for exploration in the space of electrical manipulation beyond linear bias/current ramping and pulsing.  Possibilities include the use of extra electronic contacts to the sample in order to globally or locally shift electron filling of the substrate / adsorbed structures with pre-fabricated back gates, or side gates written in situ with the STM, or implementation of lateral and vertical tip-based manipulation techniques that have been demonstrated for other systems [10].  Alternative stimuli that induce desired reactions (e.g. laser-induced, or use of additional species, dangling-bond structures, or surface features that mediate reactions or add needed reaction steps during tip-based manipulation) may also present a promising avenue for progress.

**Concluding Remarks**
Tip-induced modification of adsorbed $PH_2+H$ is an effective method to achieve single P atom incorporation into STM-patterned devices.  The major investment of both time and operator expertise makes adoption of this method highly non-trivial in a field (i.e., STM-fabricated atomic-precision devices) that already experiences a large barrier to entry.  The advances described here will go a long way to overcoming this barrier, which we consider to be a valuable proposition as tip-based manipulation greatly enhances our control over the fabrication process and allows us to achieve true single-atom precision for species of interest such as $PH_x$.  Moving forward, controlling and understanding the atom-scale details and chemistry of device fabrication one atom or molecule at a time - which is most straightforwardly achieved with STM tip-based manipulation - will be a key component of perfecting our ability to fabricate devices at the ultimate limit.

**Acknowledgements**
The author wishes to acknowledge useful discussions and technical contributions from Xiqiao Wang.  This material is based in part upon work supported by the National Science Foundation under Grant No 2240377, and by the Department of Energy Advanced Manufacturing Office Award Number DE-EE0008311.

# Section 4.3 – Alternative Precursors for Semiconducting Atomic-scale Devices
Robert. E. Butera

Laboratory for Physical Sciences, College Park, MD, 20740, United States

**Status**
Within the field of semiconducting atomic-scale devices (ASD), gaseous molecular precursors have been primarily used to deliver dopant atoms to a pre-defined location on a given substrate. In general, these precursors must satisfy a stringent set of requirements that are similar, in many ways, to those associated with selective area atomic-layer deposition (ALD) [1]. Specifically, these precursors are expected to:
1. Foster a self-limiting, selective adsorption process into a chemically distinct patterned region defined within a monolayer resist,
2. Contain a single "target" atom with easily removable ligands that neither contaminate the substrate nor otherwise interfere with subsequent processing steps,
3. Possess an experimentally accessible and theoretically identifiable reaction pathway resulting in the incorporation of a single atom into an atomically precise location, and
4. Be compatible with ultrahigh vacuum (UHV) environments.

The current choice of precursors for ASD fabrication has been largely driven by the desire to fabricate single-dopant qubits for quantum computing applications which imposes further restrictions on purity and elemental composition. In total, all the above-mentioned requirements and restrictions have been taken to imply that an appropriate precursor should be free of both carbon and oxygen, with the ideal precursor consisting of a single dopant atom fully saturated with hydrogen atom ligands. As a result, $PH_3$ has been the primary precursor explored to date.

Recent theoretical and experimental investigations of numerous alternative precursors to $PH_3$ have been pursued to realize alternative devices with alternative performance [2-7]. Most notable amongst these are the halide-based precursors [4-7] that were shown to be compatible with both hydrogen and halogen monolayer resists under standard conditions. These precursors have the added benefit of providing additional functionality to the fabrication process due to their enhanced reactivity and increased size as compared to their hydrogen-based analogues. Most importantly, halide-based precursors are readily available for most elements of interest and easily incorporated into a UHV dosing system. The field is thus primed for the exploration of new materials and processes to usher in the advent of future novel ASD designs and architectures.

**Current and Future Challenges**
The first near-term challenge is identifying the most promising element, and its associated precursor, to explore that will provide a significant technological advancement or unlock additional functionality within the operation of current device architectures. In this vein, there has been recent interest in optically active elements in silicon, such as Er, for use as quantum emitters within a distributed quantum network and other photonic devices [8]. Once the precursor is identified, the atomic-level sequence of events of adsorption, desorption, and decomposition of the precursor and its by-products must be investigated with sufficient detail to obtain a thorough understanding of the incorporation of the target element into the desired location on the target substrate.

The most significant challenge for further exploration of alternative precursors is the near-exclusive association of the ASD fabrication process with single dopant qubit devices. In particular, the necessary evil of securing funding in this field typically requires the demonstration of a functioning quantum device in a limited timeframe. This perpetuates a "follow the leader" approach replicating established processes at the expense of innovation. New methods are only proposed as mitigation strategies to be explored should the current process breakdown. As such, alternative precursor innovation is wholly dependent upon the discovery and development of novel device designs and architectures that dictate the use of a new elemental specie or that require chemical alterations of the process flow.

As an example, advanced 3D ASD designs were proposed in which the heavily doped, atomically-precise active regions of the device are extended in the vertical direction [9]. This poses a significant fabrication challenge utilizing the current approach depicted in Fig. 1a, which requires multiple rounds of processing and tip realignment over the active region for each subsequent layer. A process could be envisioned where the chemical reactivity of the precursor and substrate passivation layer are leveraged to drive the desired reactions after a single patterning process, as depicted in Fig. 1b. Realizing such a process will require advancements in precursor dosing schemes, as described in the next section.

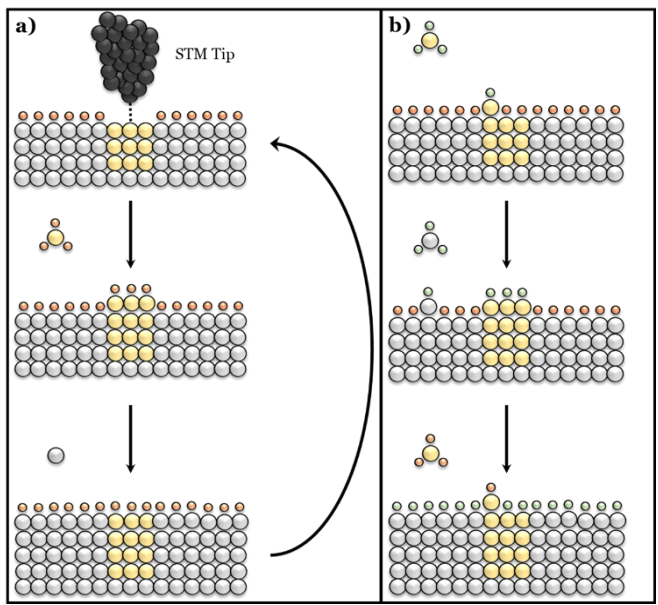

**Figure 1** a) Depiction of current ASD fabrication process for multilayer devices. The STM patterns a region in a monolayer resist, which is then exposed to a molecular precursor that selectively deposits into the patterned region. A layer of the substrate material is then deposited through physical vapor deposition, after which the region of interest must be relocated with the STM to create the pattern in the newly formed top layer, and then the process repeats. b) Depiction of a surface chemistry driven ASD process. After an initial pattern is created on the initial substrate surface, the difference in chemical reactivity of the patterned region versus the surrounding area is leveraged to drive selective deposition into either region based on the choice of chemical constituents of the alternative precursor.

**Advances in Science and Technology to Meet Challenges**
Due to the sensitivity of dopant qubits to nearby defects, a minimalist approach has been taken to limit the number of in-situ processing steps and, ultimately, minimize unintentional contamination. This results in straightforward, single-precursor dosing schemes. In contrast, ALD processes commonly employ more complex A-B-A-B… and A-B-C-A-B-C… dosing schemes, where A, B, and C are distinct precursors utilized to deposit and/or further reduce adsorbed species during the growth of thin films [1]. Such complex dosing schemes may be required to deposit certain elements that may not have a readily available molecular precursor compatible with the requirements listed above. These schemes may then enable the use of C and O containing molecules, so long as these species are easily removed by a secondary reaction step.

To facilitate these complex dosing schemes, focused efforts will be required to identify suitable precursors or investigate the design of novel precursors that meet the stringent requirements listed above and operate in sequence to drive the appropriate reaction in the desired location. As a result, there will be a strong demand for computational chemists to model not only the reactions of individual precursors with the substrate and other adsorbates, but also the reactions involved in the generation of the precursors themselves. Experimentally, process verification and validation would need to be established with traditional surface science analysis tools and techniques, such as x-ray photoelectron spectroscopy, Auger electron spectroscopy, temperature programmed desorption, and scanning tunnelling microscopy, all requiring the expert services of physical chemists and surface scientists.

To aid in the exploration of novel precursors, rapid prototyping methods should be utilized to test the capability of new families of precursors to react in the intended manner prior to implementing the costly and time-consuming safety engineering requirements that facilitate the safe usage of these mostly hazardous gaseous precursors. As an example, wet chemical methods have recently demonstrated the adsorption of dopant containing molecules to a silicon surface, which uncovered selective reactions that can be further exploited for selective deposition of alternative gaseous precursors [10].

**Concluding Remarks**
Semiconducting ASDs remain a very nascent technology explored by a relatively small community of researchers. To make significant advancements within this field, the underlying fabrication methods must find relevancy beyond devices for quantum computing applications to attract immediate interest from the research community at large. Through focused efforts and further engagements with chemists, surface scientists, and materials scientists aimed at gaining a thorough understanding of the underlying surface chemical processes associated with alternative precursors, it will be possible to engineer these precursors to not only enable targeted single atom deposition of a desired element, but also to provide additional functionality to enhance and advance the overall fabrication process. In doing so, the growth of other materials including metals, semiconductors, and dielectrics all having the requisite atomic-scale precision in both the lateral and vertical dimensions, while maintaining the chemical purity thus far demonstrated within the current processing scheme, could become reality.

**Acknowledgements**
The author would like to thank Michael Dreyer and Jon Marbey for comments and suggestions.

# Section 4.4 – Halogen resists

Tatiana V. Pavlova

Prokhorov General Physics Institute of the Russian Academy of Sciences, Vavilov str. 38, 119991 Moscow, Russia

**Status**

A monolayer resist on a Si(100) surface is used for the most precise dopant incorporation utilizing STM lithography [1]. Despite significant progress in the placement of impurities using a hydrogen monolayer, there is still potential for improvement in the dopant insertion procedure, specifically by enhancing the precision of dopant positioning, increasing the probability of dopant incorporation as a substitutional impurity, and minimizing the number of surrounding defects in the silicon epitaxial layer. The most ambitious goal is to precisely incorporate dopants into the site of a selected silicon atom. In addition, the ability to introduce various dopants is highly desirable. While a hydrogen monolayer resist remains utilized to overcome these challenges, other resists could be tried as well. The alternative resist should cover all the reactive Si dangling bonds, preventing silicon-dopant bond formation, and enabling STM lithography to create a mask. Additionally, the resist must be compatible with subsequent silicon epitaxy, segregating to the surface or desorbing at temperatures below the temperature of the epitaxy.

Currently, halogens are being considered as an alternative to hydrogen. Halogen resist is compatible with various halide dopants such as $PCl_3$ and $BCl_3$, and protects the Si(100) surface from undesired dopant incorporation even better than hydrogen [2]. Among the all halogens, chlorine is the most studied. It was originally proposed to utilize chlorine as a resist with the idea of removing not only the chlorine atom, but also the silicon atom due to the strong Si-Cl interaction [3]. STM lithography on a chlorinated Si(100) surface has already been demonstrated with the removal of individual Cl atoms [4] (Fig. 1a), the creation of one-dimensional and two-dimensional arrays of Cl vacancies [5] (Fig. 1b), and the formation of etching pits by removing Si atoms from the surface [6] (Fig. 1c). Single atom removal has also been demonstrated for a brominated Si(100) surface [4] (Fig. 1d). Other halogens are hardly suitable for the single dopants incorporation, since iodine covers only half of the Si dangling bonds and fluorine is very toxic.

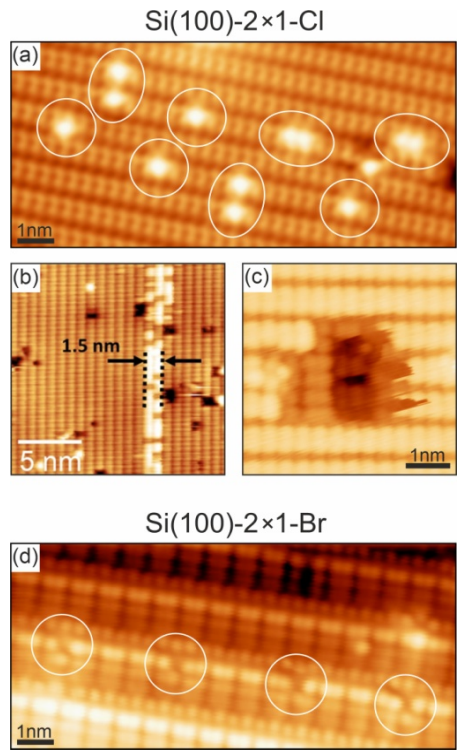

**Figure 1** STM lithography on a halogenated Si(100) surface. STM images of a chlorinated Si(100) surface with (a) single vacancies (adapted from [4] with permission), (b) the line of vacancies (adapted from [5] with permission), and (c) the etching pit with two silicon layers removed (adapted from [6] with permission). (d) STM image of single Br vacancies created on the atomic step of a brominated Si(100) surface (adapted from [4] with permission).

**Current and Future Challenges**

Although halogens are a promising resist for single-dopant incorporation, no single-atom device has yet been demonstrated utilizing them. One of the main challenges is the removal of single atoms in STM with high accuracy. In the case of a Si(100)-H surface, there is a long-lived Si-H stretching mode that allows the vibrational heating mechanism to be used to precisely remove an H atom. In the case of halogens, the lack of a long-lived vibration prevents the application of this mechanism. Nevertheless, individual halogen atoms can be removed, but with lower precision [4]. Furthermore, it would be beneficial to remove a silicon atom to introduce a dopant in its place [3]. In this case, there is no need to anneal the sample in order to replace a silicon atom with the impurity, which introduces additional inaccuracy in positioning. To date, the removal of clusters of silicon atoms from a chlorinated Si(100) surface has been demonstrated [6], but not individual silicon atoms.

To use halogens for single-dopant incorporation, it is essential to understand the reaction between the halogenated surfaces and various precursors. Initially, it is crucial to demonstrate that, for this precursor, the halogen monolayer acts as a resist by preventing dopants from bonding with silicon [2]. Then, the minimal window in the resist required for the creation of a bond between the dopant and silicon should be defined [7, 8]. Further, all possible pathways of the precursor-surface reaction must be studied in this window in halogen resist.

Finally, an important challenge is to optimise the silicon epitaxy on Si(100) with a halogen mask and dopants, taking into account two competing processes. The temperature should be high enough to promote the growth of crystalline silicon while being low enough to prevent the diffusion of precisely placed impurities (the same requirement as for hydrogen resist). During the epitaxy process, the halogens must desorb or segregate on the silicon growing surface and do not remain in the silicon lattice.

In the long term, the main challenge is to develop reliable technology capable of fabricating large arrays of precisely positioned dopants.

**Advances in Science and Technology to Meet Challenges**
To overcome the challenge of precisely creating vacancies using STM lithography on the halogenated Si(100) surface, very stable STM tips are highly desirable. Namely, a voltage pulse used to remove halogens should not alter the tip. The pulse quite often causes a change in tip configuration rather than the removal of a surface atom because the atoms at the tip are more mobile than those in the surface layer. As the structure of the tip changes, the voltage needed to remove a single atom varies in an unpredictable way. Thereby, a pulse with the same voltage either does not lead to the removal of the halogen, or results in the removal of multiple atoms at once instead of one. This challenge can be solved by creating a tip which remains stable under the action of the voltage pulses. An alternative solution is to search for a non-pulse approach to remove halogens in STM.

To employ halogens for individual dopant incorporation, it is essential to study all stages of the reaction between a precursor and a mask of halogens [2,7,8]. A good model system to study such a reaction is $PBr_3$ on Si(100)-Cl since Br and Cl can be distinguished in STM, allowing the dissociation process to be determined (Fig. 2a) [7]. To address the challenge of atomically precise incorporation, the mechanism of an impurity atom exchanging with a pre-selected Si atom must be found. It is very helpful to observe in an STM the same surface area before and after adsorption, heating, and the first step of epitaxy in order to study these surface reactions.

To create technology for single dopants incorporation using halogen resist, an epitaxial layer must be grown. According to calculations, halogen atoms are either segregated to the surface or removed as $SiCl_2$ (Fig. 2b) [9]. Experimentally, it was demonstrated that chlorine is removed from the epitaxial silicon layer, particularly efficiently at 850 K (Fig. 2c) [10]. To reduce the number of defects, further work is needed to optimize the epitaxy process and subsequent oxide formation on the partially halogenated silicon surface.

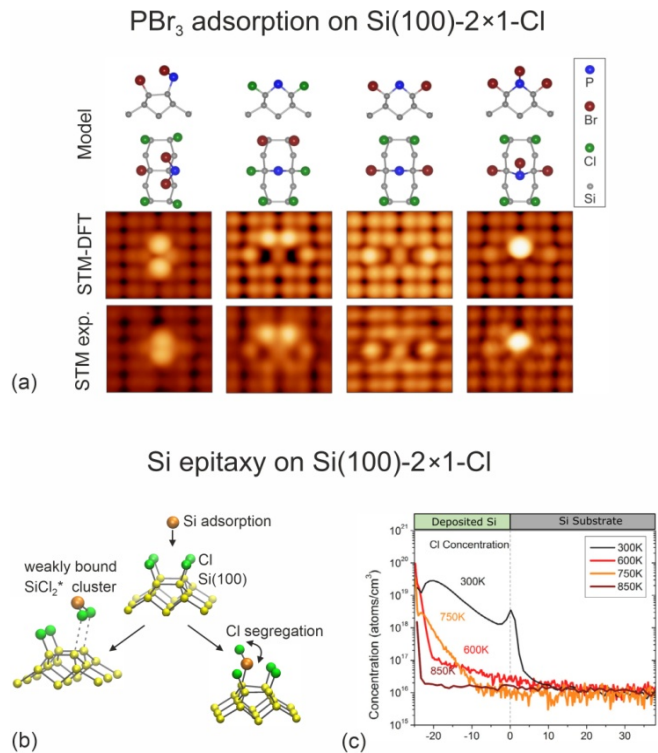

**Figure 2** (a) PBr$_3$ adsorption on a Si(100) surface through a chlorine mask with windows of two silicon atoms. Phosphorus is adsorbed as PBr$_2$ on a Si atom or as P or PBr in a bridge site on a Si dimer. Adapted from [7] with permission. (b) DFT calculation of the Si adsorption on a chlorinated Si(100) surface. The Si atom penetrates under the Cl monalayer or forms a weakly bound SiCl$_2$ fragment. Adapted from [9] with permission. (c) Epitaxy on a chlorinated Si(100) surface. At 850 K, chlorine atoms mostly do not remain in the epitaxial layer. Reproduced from [10] with permission.

**Concluding Remarks**

Halogen resist has garnered interest for silicon doping since it is compatible with halogenated precursors and is expected to provide advantages such as atomically precise doping. To successfully use halogens, a number of challenges must be resolved. The main challenge with halogen resist is the removal of single halogen atoms, which is more difficult than removing single hydrogen atoms. This problem can probably be solved by creating an extremely stable STM tip. To achieve atomically precise impurity incorporation, it is crucial to understand all steps of the precursor's interactions with a halogen mask on the Si(100) surface, including adsorption and P-Si exchange. Additionally, the epitaxy and surface oxidation processes need to be optimized. If these issues are addressed, halogen resist could be a good alternative to hydrogen for precise dopant incorporation.

**Acknowledgements**
This work was supported by Russian Science Foundation under grant No. 21-12-00299.

## Section 5.1 – Novel methods in ion implantation

A.M. Jakob[1,2], D. Spemann[3], P. Räcke[3,4], F. Schmidt-Kaler[5] and D.N. Jamieson[1,2]

[1]School of Physics, The University of Melbourne, 3010 Parkville, VIC, Australia
[2]ARC Centre for Quantum Computation and Communication Technology (CQC$^2$T)
[3]Leibniz-Institute of Surface Engineering (IOM), 04318 Leipzig, Germany
[4]Universität Leipzig, Felix Bloch Institute for Solid State Physics, Applied Quantum Systems, 04103 Leipzig, Germany
[5]QUANTUM, Institut für Physik, Johannes Gutenberg-Universität Mainz, 55128 Mainz, Germany

**Status**

Ever since the first demonstration of a silicon transistor concept, its rise to a trillion-dollar industry was paved by ion implantation technology. Enabling the reliable doping of semiconductor materials with foreign atoms to precisely tailor their local electronic properties, it shaped the progressive evolution of classical computer processors to state-of-the-art ultra-scaled nano-FinFET architectures. To date, it's scalability forms a central pillar to keep the pace with an ever-growing global demand on powerful processor technologies.

Focusing on dopants in semiconductors, the seminal Kane proposal marked a new era by adapting the revolutionary quantum computer concept to a silicon CMOS device [1]. By exploiting entangled networks of nuclear-electronic $^{31}$P donor spins for quantum information processing, it offered the silicon semiconductor industry prospects of an intuitive path towards scalable quantum processor fabrication, using established methods such as the industry standard ion implantation.

Indeed, since then, ion-implanted donors in silicon have been demonstrated to exhibit exceptionally long spin coherence times and enabled sophisticated gate readout fidelity tomography on multi-qubit processors [2] [3] [4]. These unique achievements place them at the performance apex of qubit platforms in the solid state. During the last decade, further key advances in quantum error correction and donor spin control schemes have expanded their capability portfolio decisively [5] [6]. Akin to quantum dot technology, implanted donors thus evolved to one of the most promising roadmaps towards an interconnected diversified quantum processor architecture in silicon.

However, in contrast to classical layouts, a quantum processor needs a 2D precision array, composed of millions of dopant atoms to unleash the promised computational power of a fully entangled qubit network [7] [8]. Once again, a fundamentally different device design entails a radical shift of production line requirements - urging implantation technology to deliver control over single ion precision placement at the industry scale.

Presented here are four deterministic single ion implantation methods that show promise for the controlled precision placement of single dopant atoms [9] [10] [11] [12] [13] [14] [15] [16] [17] [18] [19] [20]. However, an operational nano-electronic qubit device, based on deterministically implanted dopants has not been demonstrated yet.

**Current and Future Challenges**

The ongoing advances in multi-donor qubit control schemes and quantum error correction demonstrate the useful role of ion implantation technology in the production of devices that will form the building blocks of a large-scale quantum processor. In this respect, two criteria can be used to formulate a state-of-play performance framework for ion implantation technology – namely the need for:

(i) an ever-improving tolerance threshold, requiring ≳90% operational qubits in a network [21] [22]
(ii) an increasing tolerance to the lateral donor placement uncertainty to ≲15 nm [5]

Future challenges will largely depend on the parallel evolution of control schemes and technology, however, current challenges are connected to one important attribute of dopant qubit devices in semiconductors. Their nano electronic surface circuitry operates reliably only in a narrow dopant placement depth window of ~5-20 nm - owing to limitations of qubit tunnel coupling [23]. As a result, dopants must be implanted at ion energies of ~10-20 keV [24]. This physical constraint entails two immediate challenges:

*I.  Control over low-energy single ions:*
A high confidence in implanting single keV-ions is crucial to comply with criterion (i). This is due to additional fault sources factoring into the final qubit array fidelity, such as the placement uncertainty of a dopant and its 'activation' to an operational qubit platform. To date, this 'activation yield' varies widely anywhere between ~64% ($^{209}$Bi in Si [25]) and ~100% ($^{31}$P in Si [26] [27]). Hence, further process optimisation is required for many dopant-substrate qubit platforms. Maximising the fidelity of single ion control to near unity confidence (≳99%) at ~10-20 keV implantation energy is therefore desired.

*II.  Control over placement position:*
Spatial straggling is inherent to random ion-solid collisions upon implantation and determines the ultimate placement precision of a dopant [24].
Depth straggling can cause a dopant fraction placed outside the window for reliable qubit control and violating criterion (i).
The lateral placement uncertainty is additionally constrained by the ability of directing ions to the aspired placement site (e.g. via focusing or collimation). It superimposes to the lateral ion straggling in the solid. At present, even for the least restrictive 'flip-flop' qubit control scheme [5], the combined uncertainty cannot exceed ≲15 nm according to criterion (ii).

**Advances in Science and Technology to Meet Challenges**
During the past ten years, four single ion implantation schemes have engaged challenges I and II by adapting core components of implanter technology in fundamentally different ways. They can be categorised into two pre-implant and two post-implant detection methods.

The pre-implant detection schemes aim to register single ions prior to their implantation [13] [15]. They avoid substrate contamination by permitting only desirable dopant ions, however, they cannot ascertain the success of the actual implantation event.
Post-implant detection schemes register the ion implant event upon substrate impact [9] [10] [11] [12] [14] [17] [18] [19] [20]. They are therefore compatible with both, electrostatically focused ion beams (FIBs) and mechanical collimators (e.g. scanned nano apertures, lithography masks) for lateral dopant placement control.

The attributes of the four single ion implantation schemes are tabulated here. Their principles, scope and development status are highlighted considering challenges I and II.

| | Pre-implant Detection | |
|---|---|---|
| **Method** | **Image Charge Detector (ICD)** | **Paul Trap Gun (PTG)** |
| **Principle** | Hollow cylinder electrodes in series. An ion passing through this linear array with defined velocity, induces an alternating image charge signal | Deterministic single ion emitter concept using a series array of Paul traps. Upon ejection of a single ion at the array exit, the emitter is reloaded by sequentially |
| | with known equivalent frequency. This is extracted from noise via Lock-In amplifier. Signals below acceptance threshold are rejected, triggering ion discrimination via electric blanker field [15]. | passing on single ions between neighboured traps. Ejected ions are accelerated and focused/directed onto target via electrostatic lenses and scanning plates, respectively [13]. |
| **Setup configuration** | Situated in a modified FIB system between ion source and focusing lens assembly. | Ultra-high-vacuum chamber Paul-trap. Laser cooling and electronics adopted from trapped ion qubit technology. |
| **Primary scope** | Proof-of-principle system for industry scale dopant array engineering in arbitrary substrates. Activation/conversion yield surveys of dopants/colour centres. | Shallow precision placement of small single dopant arrays in arbitrary substrates. Conversion yield studies on Colour Centre arrays and $^{141}$Pr in YAG crystals |
| **Key Challenge I** ≳99% confidence for single 10-20 keV ions | Not yet demonstrated. Can detect bunches of ~80 elementary charges with a false positive rate of ~0.1% [28]. | ~100% demonstrated for sub-10 keV $^{40}$Ca$^+$, $^{15}$N$^+$, $^{141}$Pr$^+$ ions [13] [16] [29]. By design, the concept exhibits ~100% single ion ejection confidence. |
| **Key Challenge II** ≲15 nm placement precision | Not yet demonstrated. ~180 nm lateral ion focus shown [30]. Improvement to ~25 nm anticipated. | Demonstrated for 6 keV $^{40}$Ca$^+$ (~6 nm focus) [13]. Not yet shown for application-relevant dopants (~120 nm focus for $^{15}$N$^+$ [29], 30 nm focus for $^{141}$Pr$^+$ [16]) |
| **Outlook:** | Further improvement of signal-to-noise ratio via electrode capacitance reduction and advanced preamplifier for single ion sensitivity. Requires ion sources capable to generate high ionisation states. | Fast sequential single ion implantation for semiconductor dopant qubit platforms. Implementing alignment tools for compatibility with qubit device fabrication process. Determining gun ejection rate for scaling potential |
| **Post-implant Detection** | | |
| Method | **Ion Beam Induced Charge (IBIC)** | **Secondary Electron Detector (SED)** |
| **Principle** | Implantation substrate is configured as single ion detector via on-chip p-i-n electrodes. Upon implantation, an ion generates free electron-hole pairs ('charge'), inducing a charge signal at the reverse-biased electrodes [9] [10] [12] [14] [18] [19]. | Upon implantation into a substrate, an ion induces ~1-20 secondary electrons (subject to ion species and substrate material for 10-20 keV), leaving the sample to trigger a signal in a channel electron multiplier plate array [17]. |
| **Setup configuration** | Single-ion-detector substrate wire-bonded to a highly integrable miniaturised charge-sensitive amplifier PCB. Realised in two configurations: **C1.** Integrated to a target AFM stage of a conventional ion implanter beamline. Spatial ion collimation via cantilever nano aperture [9] [18] [20]. | Installed in close vicinity to the sample stage of commercial FIB systems to increase solid angle and improve secondary electron capture yield. |
| | C2. Integrated to an interferometrically controlled precision stage of a commercial Raith FIB system [14] [19]. | |
| **Primary scope** | Deterministic single and multi-donor qubit silicon devices. Small and meso-scale deterministic donor arrays for scalability tests. | Scalable dopant array engineering in arbitrary substrates. Enable fundamental studies on dopant/colour centre activation/conversion yield. |
| **Key Challenge I** ≳99% confidence for single 10-20 keV ions | >99.9% demonstrated for multiple donor qubit platforms in silicon: $^{31}$P (9 keV), $^{123}$Sb (18 keV) and $^{209}$Bi (20 keV) [20]. | Not yet demonstrated. ~80% single ion detection confidence for 25 keV $^{209}$Bi+ ions in silicon [17]. |
| **Key Challenge II** ≲15 nm placement precision | Not yet demonstrated. Deterministic single $^{123}$Sb donor and $^{31}$P arrays formed with 45 nm aperture (**C1**) [20]. 15 nm ion apertures fabricated but not yet tested. | Not yet demonstrated. FIB spot size of ~15nm demonstrated for 50 keV $^{209}$Bi [17]. |
| **Outlook** | Single ion detection performance tests with $^{28}$Si-enriched substrates. Combining fabrication flow of single ion detector and qubit control circuit to demonstrate deterministic silicon donor qubit device ($^{31}$P, $^{123}$Sb). | Further improving single ion detection confidence of SED. Developing fabrication flow for silicon doner qubit devices. Deterministic $^{209}$Bi arrays for EDMR device fabrication |

**Table 1** Overview of deterministic single ion implantation technologies under development.

**Concluding Remarks**

Single ion implantation shows promise to deliver deterministic single and few dopant qubit devices in near future. A maturation strategy for scalable quantum processor engineering should focus on complementary advances at two frontiers:

The Quantum Frontier that fosters enhancement of qubit control schemes towards even higher qubit loss fault tolerance (~75% yield). Of interest are architectures with more degrees of freedom for dopant qubit entanglement, such as array geometries with >4 neighboured dopants and nested arrays, allowing optional next-neighbour coupling. Hybrid architectures mediating dopant qubit entanglement via quantum dots [31] [32], cavities , waveguides [33] or superconducting circuits [34] [35] could further mitigate placement precision limitations of ion implantation. The latter three as well as micro-optical resonators offer additional maturation roadmaps for other promising qubit platforms like Colour Centres in diamond [33].

The Materials Engineering Frontier foremostly covers systematic substrate annealing surveys to improve the activation yield for near-surface placed dopants. Here, exploring laser and electron induced anneal schemes could support conventional thermal approaches. To date, single ion implantation methods require further improvement of their ion collimation/focusing technology to ~5-10 nm to preserve a lateral dopant placement precision of ≲15nm.

Mitigation strategies for improved dopant placement depth precision to ~5nm may include the shallow sub-5 nm single ion implantation at ~5-10 keV, followed by homo-epitaxial regrowth of the substrate material.

**Acknowledgements**

The University of Melbourne authors acknowledge national funding schemes from Australia (ARC CQC2T, NCRIS HIA), the USA (ARO NINES) and the UK (Royal Society Wolfson Fellowship). UA-DAAD travel scholarships are greatly appreciated.

# Section 5.2 – Atomic Fabrication by Electron Beams

Utkarsh Pratiush, Gerd Duscher and Sergei V. Kalinin

Department of Materials Science and Engineering, University of Tennessee, Knoxville, TN 37996, USA

**Status**

Electron beam induced changes in materials structure has been known in electron microscopy since the first experiments by Ernst Ruska before WWII. Traditionally considered as purely deleterious beam damage, this was one of the preponderant factors (along with resolution) driving the development of electron microscopy over decades – first towards high-voltage instruments (eighties) and then (from late nineties) towards aberration corrected (AC) instruments. The broad introduction of AC microscopes over the last decade allowed to localize beam damage to a single atomic column or chemical bond[1] – opening a new paradigm for direct atomic-level manipulation of matter.[2]

Over the last decade, this approach has been used to enable direct atomic motion,[3, 4] building homo- and multiatomic artificial molecules in 2D materials,[5] and atomic-plane sculpting of 2D and 3D materials.[6] With time, the number of materials in which atomic-beam induced transformation can be controlled on the atomic level has been increasing to include many layered and bulk materials including oxides, dichalcogenides, phosphochalcogenides, and many others.[7] These observations suggest that electron beam fabrication can become the enabling technology for areas such as nanopore fabrication for protein sequencing, molecule screening platforms for physics and biology, and particularly quantum communication, sensing, and computing devices.

To estimate the potential technological base for these developments, we note that electron microscopy is by now one of the biggest areas of instrumental research, with hundreds of high-end aberration-corrected tools worldwide and the global market size of ~$3.7 billion in 2021, projected to reach $8.3 billion by 2031. Over the last several years, many of the manufacturers started providing Python APIs for microscopes, allowing for direct deployment of machine learning based workflows to enable automated and autonomous experiments on multiple platforms worldwide. Combined with the proliferation of cloud technologies, this creates an opportunity to create an ecosystem of instruments, data repositories, and analysis codes that can vastly accelerate the discovery rate across the nano- and quantum science communities.

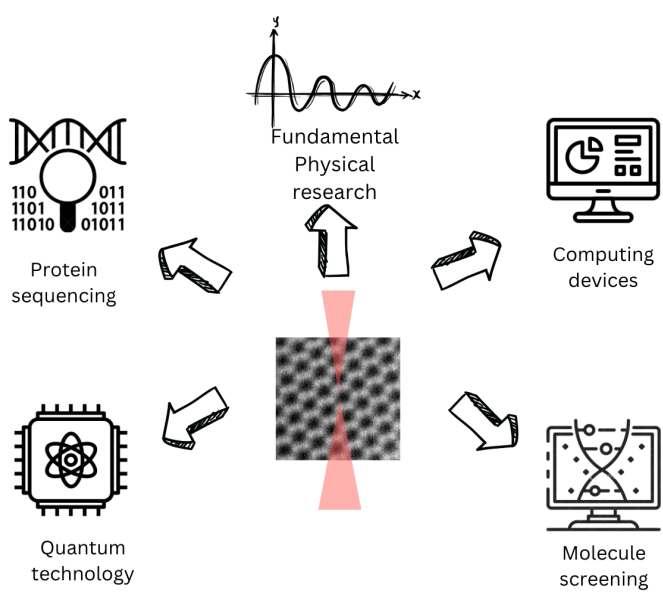

**Figure 1** Downstream applications for direct atomic fabrications via electron beams

**Current and Future Challenges**
The instrument base for electron beam atom by atom fabrication is broadly available by now in the form of thousands of platforms worldwide. The broad adoption of these methods faces three groups of problems, namely (a) limitations of human control, (b) integration between surface science and electron microscopy platforms, and (c) downstream applications and integration in current technology trees.

**Instrument control:** The electron microscopes are traditionally built as pure imaging and spectroscopy platforms using small number of established scanning modes. Until now, atomic manipulation has been performed preponderantly via direct control by human operator one beam positioning at a time based on visual feedback, and the elements of relevant know-how have been limited to only several research groups. The characteristic timescale of human-operated experiments (seconds) vastly exceeds the intrinsic latencies of the electron microscope, for which hundreds of fabrications steps per second should be possible. Similarly, human control necessarily lacks precision, reproducibility, and systematic error correction capability. While sufficient for the proof of concept, atomic scale fabrication with high precision and throughput requires moving beyond the current human control paradigm. This is particularly the case for **atomic fabrication in 3D materials** that will require the multiple tilt series to enable and control atomic motion in the non-parallel planes to achieve the full 3D control of atomic assembly.

**Integration with surface science tools:** The second key challenge integrates surface science with chemical functionalization for STEM samples, expanding the field beyond the serendipitously-present contaminant chemistries. This can involve embedding desired dopant atoms via deposition and sputtering using the electron beam from an aligned membrane, or the introduction of element sources common in surface science[8,9], including ion deposition. This intricate process allows for random dopant placement and subsequent electron beam manipulation. Similarly, chemical modification of native defects with subsequent beam manipulation can be explored using surface spectroscopies such as XPS, and diffraction techniques like LEED and SPA-LEED. Post-deposition chemical functionalization methods can be applied to confer additional stability and protection to the device, ensuring its integrity before it exits the controlled STEM environment. These preparatory steps are crucial for the subsequent fine-tuning of the device's properties and performance.

**Downstream applications:** Finally, the key technological challenge is integration between the electron beam fabrication and the downstream semiconductor technology applications. To demonstrate that the loop on the device fabrication can be closed, the atomic manipulation should be performed in the active region of the device supported on the beam-transparent membrane and interfaced with the macroscopic source, drain, and gate electrodes. Ultimately, these downstream applications will establish the practical value of this technology.

**Advances in Science and Technology to Meet Challenges**
Building electron beam fabrication for atomic scale devices requires significantly expanding the capabilities of modern scanning transmission electron microscopy in terms of beam control and rapid feedback. This includes technical challenges of the development of algorithms for rapid on-the fly image analysis and conversion of the data stream from the detectors to the individual atomic positions. Currently, this can be accomplished using the ensembled deep convolutional networks;[6] in the future robust machine vision methods that can operate under the out of distribution drift environment are required. Closely coupled to this is the problem of precise beam positioning and position refinement to compensate for inevitable non-linearities and hysteresis in beam control systems. Finally, the most crucial challenge is the deep understanding of the deterministic and probabilistic experimental mechanisms underlying beam-induced phenomena, notably the reaction rates initiated by electron beams. Machine learning, particularly contextual bandit algorithms, is poised to play a pivotal role in the early stages of this development. These algorithms can quickly adapt to changing conditions, making real-time decisions based on the current state of the system.

For atomic fabrication, where atomic interactions and local energy states are critical, myopic optimization[10] can help ensure each atom's placement is optimal for immediate stability. As the complexity of tasks escalates myopic optimizations being short sighted fails, Reinforcement learning[11] (RL) can adjust parameters dynamically, accounting for long-term growth stability rather than moment-to-moment optimization. RL will become indispensable, allowing for sequential, informed decision-making that refines the beam's behaviour to culminate in the desired atomic configurations. The implementation of RL will require a seamless, rapid exchange of information between the microscope's control systems and edge electronics, potentially enhanced by cloud computing. Moreover, the capacity to dynamically alter beam energy will empower the system to selectively activate various physical mechanisms, honing the precision of fabrication processes. This sophisticated orchestration of hardware and advanced machine learning techniques could revolutionize our ability to manipulate matter at the most fundamental level, driving innovation across nanotechnology and materials science.

The important complement to the experiment based phenomenological knowledge will be theory capable of the prediction of electron beam effects on solids. Notably, due to their high energy this requires modelling beyond Born-Oppenheimer approximation.[12] These can include nonadiabatic MD simulations and mixed quantum-classical nonadiabatic MD methods to account for the coupled electronic and nuclear evolution.

Finally, subsequent growth of the field can benefit from the development of the dual-beam machines, with low energy beams used for material imaging with minimal beam induced damage, and high energy beams for beam-induced modifications.

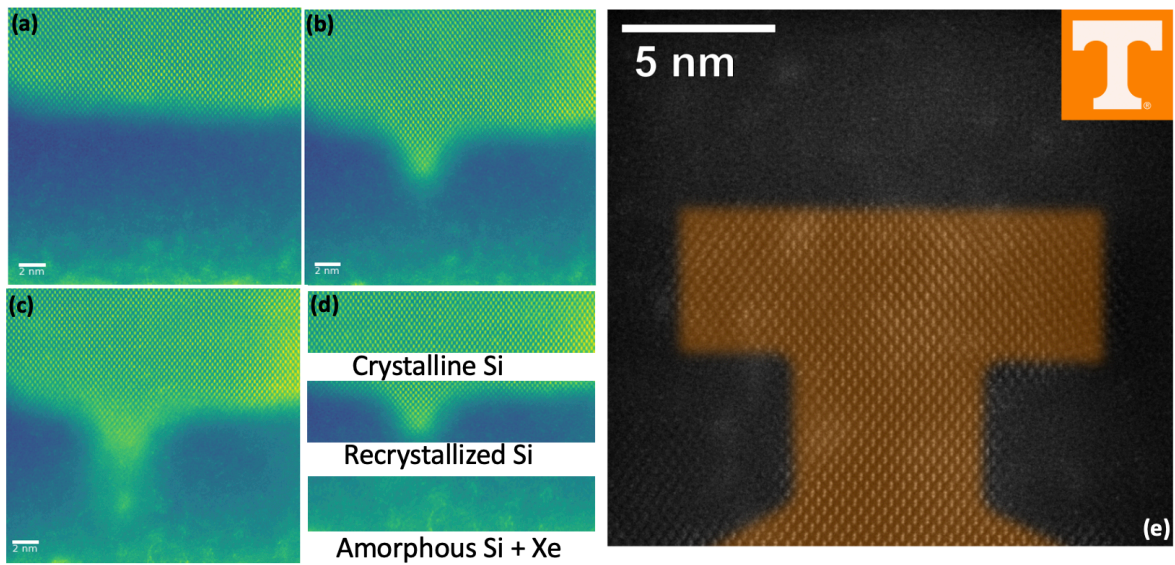

**Figure 2** Stages of Silicon (Si) recrystallization. (a) Initial stage with amorphous Si, (b) intermediate stage showing partial crystallization, and (c) final stage with fully crystallized Si structure. (d) summarizing the stages of recrystallization.", with (e) Illustration of Si recrystallizing to form a specific shape labelled **'T'[Coloured post-acquisition to orange to match University logo].**

**Concluding Remarks**

Electron beam fabrication is as a third paradigm for direct matter assembly atom by atom, crucial for domains requiring atomically precise fabrication. This technique promises value creation from a small number of components with a minimal technological footprint. In quantum computing, sensing, and communications, the precise fabrication of defect centres is essential for device functionality. For biology and medicine, it enables the design of highly specific atomic configurations such as nanopores for protein sequencing for precision medicine and drug testing. Finally, for more speculative applications, e-beam fabrications allows the creation of vital electronics on a small scale in remote extraterrestrial settings, offering an alternative to transporting bulky equipment from Earth. For fundamental physics and prototyping, atomic control is invaluable for building new molecular structures and exploring their functional properties.

## Section 5.3 – Quantum device fabrication with EUV lithography

Dimitrios Kazazis[1], Procopios Constantinou[1], Gabriel Aeppli[1,2,3,4] and Yasin Ekinci[1]

[1] Paul Scherrer Institute, 5232 Villigen, Switzerland
[2] Institute of Physics, Ecole Polytechnique Fédérale de Lausanne (EPFL), 1015 Lausanne, Switzerland
[3] Department of Physics, ETH Zürich, 8093 Zürich, Switzerland
[4] Quantum Center, Eidgenössische Technische Hochschule Zürich (ETHZ), 8093 Zürich, Switzerland

**Status**

Quantum computing is thought to be the next paradigm to transform information technology. Although there are various approaches to quantum computing, silicon-based quantum devices stand out, as they are, in principle, compatible with standard CMOS [1]. This field is rapidly evolving, driven by advances in the precise atomic-scale patterning of dopant atoms [2, 3]. At present, laboratory-scale quantum-electronic devices are created using scanning tunnelling microscopy (STM) based hydrogen desorption lithography. With this method, single dopant atoms like arsenic, boron, or phosphorus are precisely incorporated into the silicon lattice, enabling the fabrication of many silicon-based quantum devices [4, 5]. However, despite the unmatched atomic-scale resolution of STM lithography, it is a very slow and serial patterning method. Therefore, scaling up STM lithography for the fabrication of wafer-scale, commercial quantum-based electronics is not a viable approach.

The fabrication of commercially viable silicon quantum devices must rely on the high-volume manufacturing methods of CMOS technology. Photolithography is a high throughput patterning method and the workhorse of modern semiconductor high volume manufacturing (HVM). With photolithography, the aerial image created by illuminating a patterned photomask is recorded in special photoresist materials and subsequently transferred to the silicon substrate. Cutting-edge semiconductor devices are manufactured using extreme ultraviolet (EUV) lithography at a wavelength of 13.5 nm, which has been the state-of-the art in photon-based lithographic patterning since its introduction in HVM in 2019 EUV lithography is expected to reach single digit nanometre half-pitch resolution in the coming decades.

Recently, the field of EUV-based fabrication of quantum devices has been opened by two results from the Swiss Light Source (SLS). First, patterning of partially hydrogen-terminated silicon by EUV light was demonstrated without the use of a photoresist [6]. In this work, the EUV photons were generated by the SLS electron synchrotron rather than the Sn plasma of commercial scanners, to create a high-resolution lines/space pattern on silicon. The EUV light promoted the growth of a dense oxide in the exposed regions. Figure 1 shows an SEM image of 75 nm half pitch (HP) lines in silicon patterned in this way; the oxide growth mechanism was studied with X-ray photoelectron spectroscopy (XPS). More recently, Constantinou et al. demonstrated that hydrogen can be desorbed from a monohydride Si(001):H surface (the starting surface for STM-based patterning) under ultrahigh vacuum (UHV) conditions by exposure to high-intensity, synchrotron EUV light. The H desorption was quantified using STM, XPS, and X-ray photoemission electron microscopy (XPEEM) [7].

**Current and Future Challenges**

Demonstrating that H-depassivation lithography can be achieved by use of EUV photons enables, in principle, EUV technology to be used for upscaling the fabrication of Si-based quantum devices. State-of-the art EUV lithography is expected to reach 8 nm resolution in the near future and further improvements are foreseen in long term. This resolution is not comparable with the atomic precision of STM lithography. However, EUV lithography has the advantage of high throughput. Combining these two techniques in the same infrastructure can pave the way to systems where atomically precise devices to store and manipulate qubits are fabricated by STM lithography and larger-scale dopant arrays, gates, and interconnects by EUV lithography. Figure 2 illustrates the corresponding process flow.

At the moment, there is no system that combines high resolution EUV patterning with STM lithography and the necessary infrastructure around it, such as gas doping and molecular beam epitaxy (MBE) of Si. There exist several research tools based at synchrotron facilities dedicated to EUV lithography [8] and interference lithography [9]. These systems with resolution below 10 nm, have been used extensively to develop photoresist materials for EUV lithography. Therefore, they have been exposed to hydrocarbon and other contaminants and cannot readily operate under UHV conditions and be connected to an STM system. They can, in principle, be adapted to work under UHV conditions and showcase the feasibility of combining EUV and STM lithography to develop Si-based quantum electronics at a research laboratory level.

Considering the integration of STM H-depassivation lithography with EUV lithography at an industrial level, there is currently no full compatibility of the two techniques for HVM. EUV scanners work in low hydrogen pressures to prevent contamination from Sn that is used as a target element in the EUV sources and ensure the carbon cleaning of the optics, which can be contaminated by the breakdown of hydrocarbons in photoresist materials. On the contrary, STM H-depassivation systems are very sensitive to atomic imperfections and impurities and therefore require UHV conditions to ensure clean and atomically flat Si:H surfaces. Therefore, a customized version of the current tools, dedicated for the processes illustrated by Fig. 2 is needed.

**Advances in Science and Technology to Meet Challenges**
Single or few qubit array operation has been demonstrated with STM H-depassivation lithography, despite the challenges in achieving this in academic environments. Although EUV H-depassivation lithography has recently become a reality [7], there has been no demonstration of a quantum device fabricated using this technology. Significant effort should, therefore, go towards the creation of well-controlled, well-characterized and reproducible δ-layers of As or P in Si by EUV H-depassivation lithography and subsequently towards the development of few dopant quantum devices. Reliable and reproducible qubits require enhanced coherence times, short gate times, high fidelity gates, long-range qubit coupling and low device-to-device variability.

Towards achieving this ambitious goal, it is imperative to establish laboratory-scale ecosystems that facilitate large-scale EUV patterning of quantum devices. These ecosystems should seamlessly integrate EUV and STM lithography techniques, alongside essential processes such as surface preparation, gas doping, and MBE for encapsulation. It is crucial that, throughout the entire fabrication process, the wafers remain under UHV conditions to ensure optimal outcomes.

One of the primary challenges encountered in this domain is enhancing the resolution of patterns created using EUV lithography. To date, hydrogen depassivation has been the only process demonstrated, which is subsequently followed by steps such as doping, annealing, and encapsulation. However, there are significant gaps in our understanding of the achievable resolution, surface roughness, and the reproducibility of the features created through this method.

Another hurdle is achieving precise overlay in the integration of EUV and STM lithographies. This is essential for ensuring the compatibility and seamless integration of the patterns created by these two distinct processes. We are optimistic that the alignment schemes already developed for the existing STM lithography technology will be applicable here. [10]

In addition, throughput requirements might be demanding, since EUV-induced H-depassivation is a relatively inefficient process requiring doses up to 18000 $J/cm^2$ [7]. Modern EUV lithography sources come with a power of 300 W, which is reduced to roughly 6 W at the wafer due to the multiple reflections in the illumination and projection systems of an EUV scanner. Future sources are expected to reach powers up to 800 W [11]. With such power levels, the H-depassivation exposures can be on the order of several minutes per chip. This is a very low throughput compared to classical processor fabrication. Quantum devices, however, require significantly fewer qubits to outperform classical devices and therefore longer exposure times can be tolerated.

**Concluding Remarks**
Si-based quantum computing has the significant advantage of being compatible with standard Si CMOS processing. Thus, both quantum and classical components can be fabricated on the same chip in a single foundry. Although research on Si-based quantum devices has increased over the past few years, only few qubit systems have been demonstrated so far with large device-to-device variability. Scaling up the manufacturing of quantum electronics cannot rely solely on a serial patterning method such as STM lithography, despite its unprecedented atomic-scale precision. Modern electronic chips comprise billions of transistors and a key enabler has been the introduction of EUV technology in HVM in 2019. Showcasing H-depassivation lithography with EUV light is a breakthrough that enables the development of wafer-scale quantum devices. This is an exciting new topic and the fabrication of useful Si quantum devices by EUV or a mix of EUV and STM lithography at a research laboratory scale represents the next milestone towards HVM of single atom-based quantum processors.

**Acknowledgements**
We thank L.-T. Tseng, P. Karadan, T. J. Z. Stock, N. J. Curson, S. R. Schofield, M. Muntwiler and C. A. F. Vaz for the collaborations leading to the results described. GA acknowledges funding from European Research Council HERO Synergy grant SYG-18 810451 and PC was funded by the Microsoft Corporation and EPSRC during the performance of the work reviewed here.

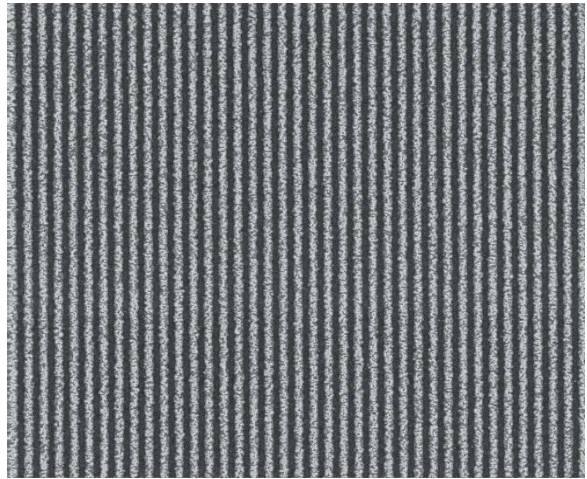

**Figure 1** 75 nm half pitch lines patterned without the use of a photoresist, etched 31 nm into Si. Adapted from [6].

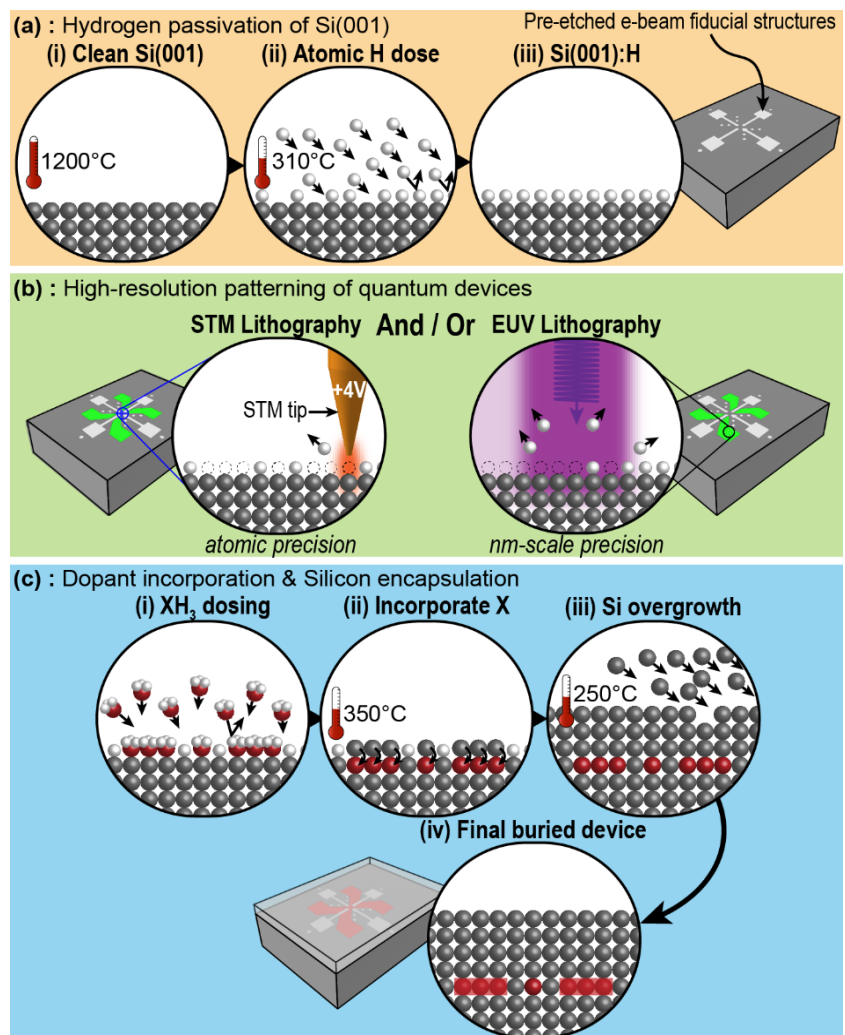

**Figure 2** Illustration of a laboratory process flow for the creation of large-scale silicon-based quantum devices in UHV. The fabrication process can be divided into three main stages: (a) The process begins with preparing a clean silicon surface, followed by the atomic hydrogen passivation of Si(001). (b) STM or EUV hydrogen-desorption lithography for patterning large-scale contacts and / or nm-scale quantum devices. (c) Incorporation of dopants and the encapsulation of silicon. After these steps, the sample can be removed from UHV conditions and exposed to ambient conditions. Standard cleanroom processing techniques are then employed to establish vertical electrical connections with the buried device.

# Section 5.4 - Scaling up hydrogen depassivation lithography
James H.G. Owen[1], Emma Fowler[2] and S. O. Reza Moheimani[2]

[1]Zyvex Labs, 1301 N. Plano Road, Richardson Texas, 75081 USA
[2]University of Texas at Dallas, Texas, United States of America

**Status**
With the ability to remove single atoms from the surface, and place single atoms onto the surface, HDL is at the limit of precision as a patterning technique. Patterning then becomes a digital process, as patterns are measured in atoms, not an analog unit such as nm. For dopant-based quantum devices, such as qubits and 2D quantum metamaterials[1], HDL remains the only way to place dopant atoms with the required precision, with current processes able to achieve placement precision of ± 1 atomic spacing. However, here its utility to limited to one material system, silicon, and a few dopant species, although more are being pursued.

In the wider semiconductor device industry, conventional lithography processes, such as e-beam and EUV lithography are reaching their limits at around 15 nm linewidths, with no viable path to shrink their linewidths any further. For further reduction in device scales, alternative lithography techniques are needed. Recently, Canon Nanotechnology has suggested that nanoimprint template lithography may be a feasible replacement for EUV. Zyvex Labs has demonstrated the possibility of creating a nanoimprint template with a linewidth as small as 6 nm by using selective Atomic Layer Deposition (ALD) of $TiO_2$ as a hard mask, followed by Reactive Ion Etching (RIE) of the silicon[2]. This suggests that HDL may be able to have an impact on the wider semiconductor industry, producing nanoimprint masks with linewidth less than 10 nm. Step and flash processes can then quickly make multiple copies of the HDL template.

However, both for large-scale production of dopant-based quantum devices, and the broader application of nanoimprint template production with a minimum feature size of a few nm, the throughput of HDL needs to be improved by several orders of magnitude while maintaining the precision. Currently, for the Atomically Precise (AP) mode, the linear tip speed is 10-20 nm/s, giving a write speed of ~100 atoms/s. In the higher-voltage Field Emission (FE) mode, where the linewidth is several dimer rows and the required dose is smaller, throughput of around 1000 atoms/s is possible, but the atomic precision is lost, and the line edge roughness is significant.

**Current and Future Challenges**
There are several different research issues that define the challenge of increasing HDL throughput while maintaining precision and reliability.

The primary issue continues to be the structure of the tip, which will need to be more stable to have a reasonable lifetime for lithography applications. Most STM users use Pt/Ir or W tips, as they have since the invention of the STM. In-vacuo sharpening techniques, such as field evaporation and field-dependent sputter sharpening show improvements in tip performance and longevity. However, even at room temperature, metal atoms on a tip can be expected to be mobile under STM conditions. Robert Wolkow[3] has developed a process of dissolving nitrogen into a W tip to make the structure more rigid and less mobile, while alternative materials such as highly doped GaN[4] have shown promise as more stable tips.

Achieving higher write speeds may be realized by exploring the limits of the current process. Joe Lyding has demonstrated single dimer row writing at higher speeds using tunnel currents up to 90 nA. However, with a tip speed approaching 1 μm/s and a tip-sample gap less than 1 nm, the standard PI control loop is likely to insufficient to prevent tip crashes. Adaptively tuning the PI controller parameters from local barrier height estimation has been shown to be effective [5]. In the long run, a superior control methodology will need to be adopted.

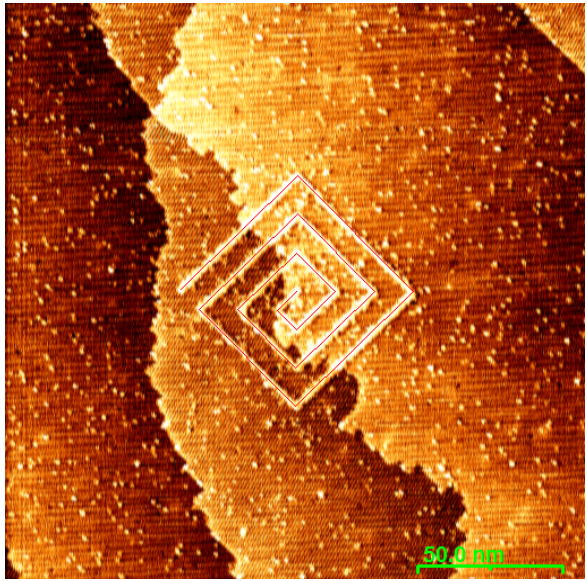

**Figure 1** STM image of an HDL spiral pattern taken at 2 µm/s using a hybrid STM, which comprises a MEMS z-actuator with the Omicron VT STM piezo tube performing xy scanning. This is 10x faster than our standard scan speed with a standard piezo-tube STM.

An alternative way to reach higher throughput is to move from a single piezo-driven tip scanning over a ± 10 µm scan field to an array of parallel MEMS-actuated tips. The mass of the piezo tube and STM tip assembly limits the z-axis bandwidth to 1 kHz; MEMS nanopositioners have the potential to increase the bandwidth and therefore scanning speed tenfold, without restricting the range of motion.[6] Second, MEMS actuators do not suffer from the creep and hysteresis present in piezo actuators. Finally, the footprint of a MEMS actuator is much smaller for the same scan range. An example of a hybrid STM, with a MEMS actuator controlling the z motion, and a piezo scanning in xy is shown in Figure 1. An array of parallel z-motion actuators that move in unison to perform raster scan lithography will have a much smaller footprint than an array of independent xyz-motion piezo actuators, although the latter may be required to maintain atomic resolution patterning.

The engineering required to assemble, drive, and control a large array of tips remains a significant challenge. Each tip will require a tunnelling control loop, control signals, and have activation voltages around 100 V. Furthermore, an automated image recognition capability will be required so that a tip can scan an area both before patterning to confirm the desired location of a pattern relative to other patterns, and after patterning, to confirm that the pattern has been written accurately.

**Advances in Science and Technology to Meet Challenges**
Single MEMS tips made of Pt, W, and Si mounted onto a MEMS actuator have demonstrated atomic resolution imaging on par with a W tip on a piezo actuator [7,8]. The MEMS actuator is mounted on a standard Omicron tip holder, as shown in Fig.2. The third leg, usually used for QPlus sensing, is instead used to deliver the z voltage signal to the MEMS actuator. In hybrid mode, the MEMS actuator handles the z motion, while the piezo tube performs xy scanning. The Pt and W (and GaN) are welded onto the MEMS actuator using Focused Ion Beam (FIB). This is not a process which lends itself to mass production. Etching the Si of the actuator into a sharp tip is an alternative route, which may be more scalable. Further study of different materials to produce a tip which remains stable and sharp over long periods of patterning will be required.

More advanced control loops are being developed. The Model Predictive Control Design methodology uses a theoretical model of the tip and sample system to predict how the control situation will change, for example as a H atom is removed, which changes the local electronic structure dramatically, and adapt pre-emptively. In order to build such a model for HDL, a much deeper understanding of the physics of the HDL process is necessary. Details such as the spatial distribution of the tunnel current, which would affect the patterning linewidth and the tip position tolerance, are currently unknown. Having demonstrated the potential of MEMS actuators, the next step is to scale up from one tip to two tips and beyond. Once a small array of perhaps 4 or 9 MEMS tips has been developed, this module can then be replicated to scale up to large numbers of tips operating in parallel.

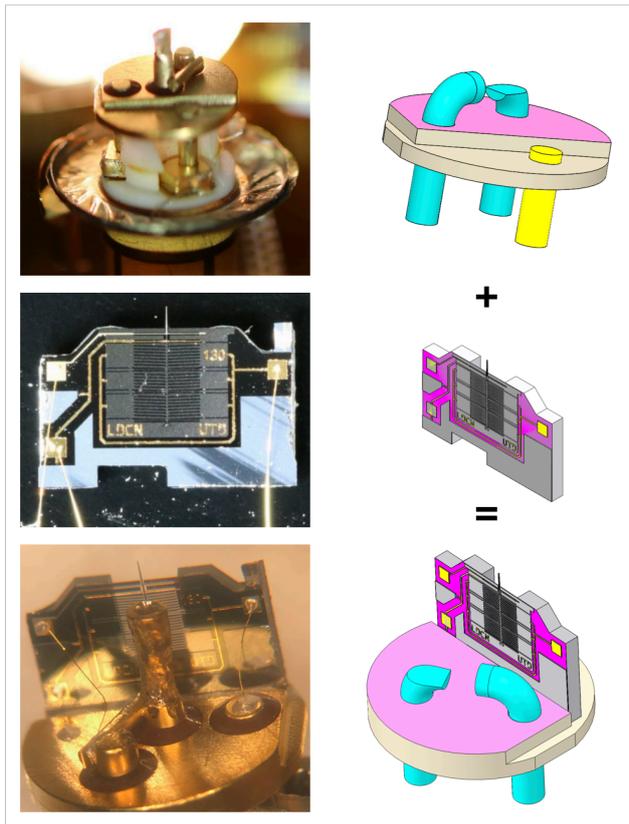

**Figure 2** The hybrid STM, so-called 'zPlus', which comprises a MEMS z-actuator mounted on an Omicron tip holder. The 3$^{rd}$ signal leg on the tip holder (normally used for Qplus) is used to provide the MEMS actuation signal, and the Omicron VT STM piezo tube performs the xy scanning.

**Concluding Remarks**

Just as optical microscopes and SEMs have had to become far more sophisticated to become optical and e-beam lithography tools, so STMs, which have barely changed in their fundamental design since their invention in 1982, will need dramatic re-engineering to become viable lithography tools for Atomically Precise Manufacturing. Innovation in tip stability, higher writing speeds, faster actuator response and parallelism via MEMS design, and advanced controller technology are promising areas of improvement to boost HDL throughput. If fabrication of nanoimprint templates using HDL is to be a feasible replacement for EUV and e-beam lithography, significant work to develop ALD and RIE processes which can produce sub-10 nm feature size with high precision, will also be crucial.

**Acknowledgements**
Parts of this work have been supported by DOE, DARPA, and the ARO.

# Section 5.5 – Digital atomic-scale fabrication
John N. Randall

Zyvex Labs, 1301 N. Plano Road, Richardson Texas, 75081 USA

**Status**
As our nanofabrication technology moves into the few nm and even sub-nm regime, more accurate fabrication is required to control and exploit the emerging properties of quantum physics. High accuracy fabrication at the nanoscale is difficult and expensive because our nanofabrication technologies are largely analog in nature. While they do exploit the chemical nature of atoms and molecules, they do not take advantage of the invariant size of atoms and molecules to control the size and position of what they are constructing.

There is an opportunity to have a $2^{nd}$ digital revolution, this time in fabrication, that will succeed for many of the same reasons that Digital Information Technology has replaced our analog information technology. Moore's Law is the poster child of an exponential manufacturing trend of improving manufacturing precision and accuracy which has driven, and limited, technological progress. As processor performance has plateaued and manufacturing costs have grown to the point where there are only a small number of companies that can participate, it is also becoming a harbinger for the end of the larger exponential manufacturing trend which is limited by the quantized nature of matter. It is time to consider a new, digital, direction in nanofabrication. In the longer run, an "Inverse Moore's Law" is realized [1,2].

Moore's Law, started at the microscale and by improving manufacturing precision permitted smaller features to realize more devices on chips that have not changed appreciably in size. The Inverse Moore's Law will start by achieving absolute/atomic precision in fabrication at the nanoscale and then <u>maintain atomic precision</u> while scaling up products over time to the macroscale. This will most likely be done by 3D mechanically realized digital atomic-scale additive manufacturing (mechanosynthesis). This process, utilizing single or multiple elements, will create atomically precise sub-components. These nanoscale parts can be hierarchically assembled to larger and larger physical scales while maintaining atomic precision using heterogeneous integration to create materials and systems with extraordinary properties and capabilities.

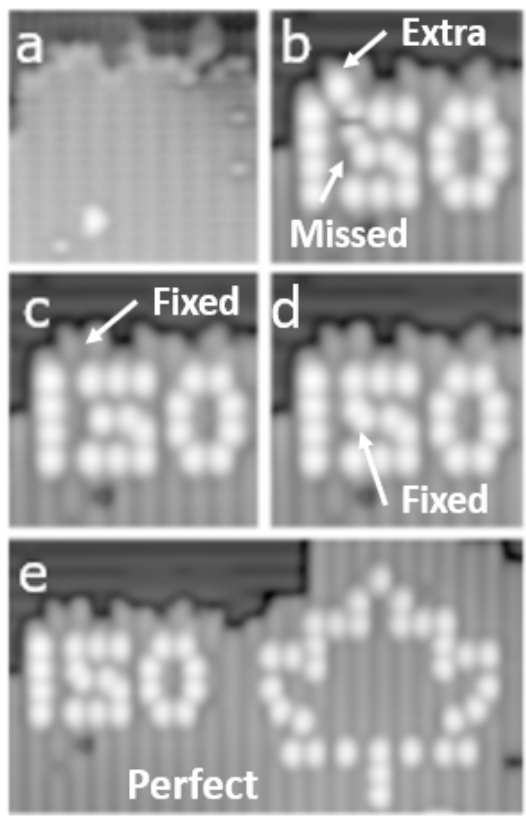

It has been proposed [1,2] that Digital Atomic-Scale Fabrication be defined as:
- Binary functions are making and breaking of chemical bonds.
- There is a spatial address grid that defines with atomic precision where bonds are made and broken.
- There are error detection and correction processes.

Two examples of fabrication technologies that meet this definition are Biology and Hydrogen Depassivation Lithography (HDL) [3]. Biology makes and breaks chemical bonds using local address grids (e.g. bases along a single strand of DNA) and does error detection and correction with polymerases. Figure 1 demonstrates that HDL satisfies all three criteria. Many other nanofabrication techniques could become "Digitized" by using techniques already developed by Digital IT.

**Figure 1** (from reference 3– reproduced with permission) demonstrates that HDL satisfies all three criteria. Si-H bonds are broken (bright spots in image), **a)** using the surface Si lattice as an address grid, **b)** errors detected by imaging, an extraneous H depassivation and missing removal of an H atom, **c)** replacing a H atom, **d)** removing an H atom, and **e)** the resulting perfect pattern.

**Current and Future Challenges**
- **Atomic precision patterning:** Hydrogen depassivation lithography (HDL) has demonstrated atomic resolution and as a digital process is capable of perfect patterning with error detection and correction [3,4]. The problem is that Tennant's Law [5] suggests that a serial writing tool with 0.768nm resolution will require >200 hours to write a square micron. Massive parallelism and significantly increased speed of a single tip will be required to create a useful manufacturing tool.
- **Atomic Layer: Deposition, Etching, and Epitaxy** – can create atomically precise thickness of thin films on surfaces, but needs a wider range of materials, uniform nucleation and to include selectivity. Error detection and correction processes need to be developed.
- **Mechanosynthesis** [6]: Directed self-assembly (lithography and selective deposition) has a significant advantage in parallel processing, but often is limited to creating materials and structures permitted by thermodynamics. In principle, the ability to mechanically position atoms and/or molecules, so they bind exactly where you want them to, is the path to a matter compiler that would create engineered structures and materials with truly remarkable properties. However, the throughput issues of fabricating Atom by Atom, are daunting.
- **Creating 3D atomically precise feedstock parts** that can be assembled into larger AP structures/systems has been referred to as "The Inverse Moore's Law" [1,2]. As with any heterogeneous integration, both the assembly mechanics and the binding technology would need to be developed.

**Advances in Science and Technology to Meet Challenges**

- **Atomic precision patterning**: An excellent description of the required advances in HDL is covered in section 4.4 "Scaling up hydrogen de-passivation lithography". These advances will increase its productivity by several orders of magnitude while retaining its sub-nm precision and digital nature.

- **Atomic Layer: Deposition, Epitaxy and Etching:** The deposition and epitaxy processes should be selective, expand their range of materials to include as many high-quality dielectrics, semiconductors, and metals as required for particular fabrication tasks, or to simply expand fabrication possibilities. ALL the deposition cycles, including the initial cycles, should be fully saturated so that isotropic growth does not create line edge roughness. Ideally the conformality of the deposition should be variable to suit the fabrication tasks. Similarly, for Atomic Layer Etching, adjusting the anisotropy of the etching would be desirable.

- **Mechanosynthesis**: The nanomechatronics that will deliver atoms and/or molecules to their desired binding sites will need something like a +/- 0.1nm positional accuracy in at least 3 degrees-of-freedom [7]. Binding sites on the assembly manipulators for atoms/molecules (parts) will need to capture the desired part, move to within capture range of the desired binding site and release it (to a stronger binding force or weaken its binding force) to allow the part to bind. There are significant power and control issues to be overcome and a high rate of assembly to be achieved mainly by parallelism[8,9]. If the Mechanosynthesis result is something like a 10nm cube, the process could be completed relatively quickly. Early demonstrations of Mechanosynthesis with even macroscale instrumentation will lead to MEMS assemblers, followed by NEMS assemblers and eventually by massively parallel nanoscale mechatronics built by mechanosynthesis.

- **Assembly of the atomically precise parts hierarchically** into larger and larger 3D atomically precise parts should be possible with suitably accurate mechatronics and possibly self-centring connectors or cold welding (Figure 2) which has been demonstrated at the nanoscale [10]. The C and D sections of Figure 2 suggests there is tolerance in assembly that preserves atomic precision. Cold welding is essentially mechanosynthesis on a larger scale. The challenges will be finding connection technology that will be appropriate for the materials used and preserve Atomic Precision.

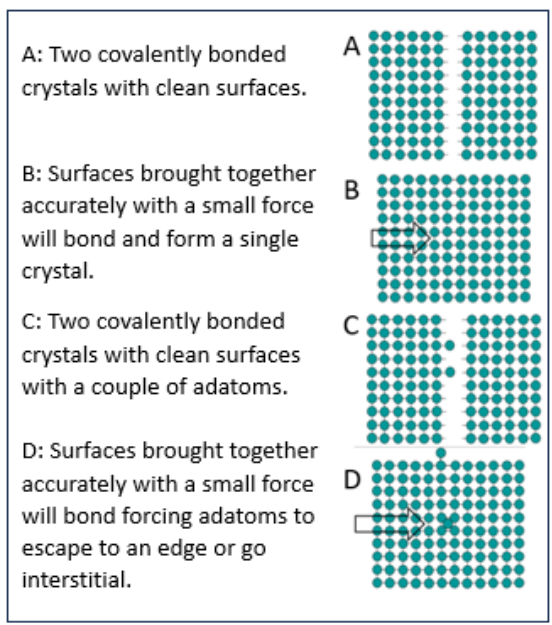

**Figure 2** A and B sections show mechanosynthesis assembly that preserves atomic precision. C and D sections suggest there is tolerance to Atomic-Scale defects that preserves atomic precision.

**Concluding Remarks**

Already, Digital Atomic-Scale fabrication has started in some 2D analog quantum simulation research [11]. The advantages in going digital have been made abundantly clear in information technology. As our nanofabrication techniques start to deal with the physically quantised states of matter, the control of dimensions of our devices and structured materials will be far easier with a digital approach. The argument against this revolution is one of manufacturing throughput. However, as we develop the ability to make atomically precise MEMS with our currently macroscale manufacturing tools, highly parallel MEMS manufacturing tools with improved throughput will produce nano electro-mechanical systems resulting in another massive increase in productivity to produce moles of nanoscale components. These become the starting point of hierarchical assembly that will eventually produce macroscale atomically precise materials and systems with unprecedented performance.

**Acknowledgements**

The author is greatly indebted to all Zyvex Labs employees over the last 10 years, particularly to James R. von Ehr, James H.G. Owen, Ehud Fuchs, Joshua Ballard, Robin Santini, Scott Schmucker, and Rahul Saini. In addition, Reza Moheimani, Yves Chabal, Joseph Lyding, Richard Silver, Shashank Misra, and Michelle Simmons have provided inspiration and valuable discussion and guidance. This work would not have been possible without significant support from DARPA, ARO, and the DOE.

## Section 5.6 – Fabrication processes and device integration

Shashank Misra, Jeffrey Ivie, Christopher R. Allemang, Evan M. Anderson, Ezra Bussmann, Quinn Campbell, Xujiao Gao, Tzu-Ming Lu and Scott W. Schmucker

Sandia National Laboratories, 1515 Eubank Blvd. SE, Albuquerque, New Mexico 87185

**Status**

Atomic precision advanced manufacturing (APAM) enables the site-specific chemical incorporation of atomic species at the surface of silicon, most famously to change its electronic properties. Careful work has established the surface chemistry of dopant precursor molecules like $PH_3$, $B_2H_6$, $BCl_3$, and $AsH_3$, through experiment and *ab initio* calculations, and the material science to preserve the placement of these dopants while encapsulating with a protective layer of epitaxial silicon. An ultra-high density of ionized dopants produce a potential strong enough to confine electrons, or holes, in 2D, 1D, and 0D and has been used to create various quantum or qubit based systems [1]. Advancing process compatibility between APAM and conventional device components is opening the door to more sophisticated electronic *devices*. In addition, the demonstration of both heterogeneous [2] and direct integration [3] open the door to more sophisticated *systems*.

Enough of the pieces required to have application impact beyond quantum computing are in place to suggest an exciting future for APAM. For example, both the potentials and carriers in highly donor-doped layers are quite different than silicon doped using ion implantation or conventional epitaxy [4], having a different band gap and carrier mass. This should enable control over quantum confinement and tunnelling in a way reminiscent of, but different than, alloy-based heterostructures. The recent discovery of a low-temperature acceptor precursor [5] opens the door to complementary transistors for digital systems, and bipolar devices for signal amplification. Benefits from such strong acceptor doping include superconducting germanium which, with APAM patterning, may enable better signal-to-noise in superconducting devices. The future discovery of chemical pathways incorporating optically active defects, or controlling the degree of local order/disorder in group-IV alloys to produce a direct band gap [6], can unlock opportunities for direct in-silicon optoelectronics. These opportunities all leverage novel material properties enabled by APAM processing rather than atomic precision patterning. To evaluate which of the aspirational application spaces in Figure 1 have a solid technical basis will require a mix of basic material science and device development.

# APAM Fundamentally Alters Semiconductor Electronic and Optical Properties

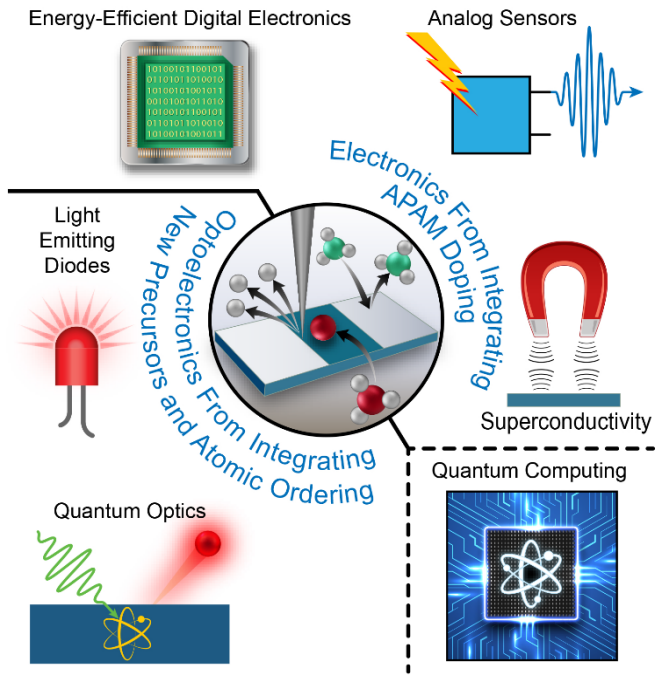

**Figure 1** APAM surface chemistry and quantum effects at the atomic scale may enable numerous applications after integration with fabrication processes.

**Current and Future Challenges**

While most of the device work in APAM has focused on the well-understood donor, phosphorus, the application spaces of Figure 1 require additional precursor discovery and characterization of the resultant material. Specifically, the next steps to determine whether there is an APAM-based advantage for optoelectronic applications is to understand the surface chemistry of many candidate chemistries and develop optical characterization of buried 2D sheets of different defects. Unlike dopants, implanted defects like chalcogens, rare earths, and various complexes cannot be deterministically placed into optically active configurations [7], where developing new precursors for APAM would enable better understanding and manipulation of these defects . Meanwhile, much of the current understanding of dopant-based APAM layers is based on electrical characterization, while we anticipate a need to better understand optical characterization. For both optically active defects and acceptors, fulfilling the need to understand the band structure will require challenging synchrotron-based photoemission experiments on *in-situ* grown films.

Developing any application will require a pivot away from one-off demonstrations and towards systematic understanding. APAM-based materials will enhance a specific aspect of a device, but whether an application advantage can be expected will require proving out a fabrication workflow that shows APAM can be integrated into a complete device, and projecting performance using device modelling that incorporates state-of-the-art non-APAM components. The former will require the consistency of the APAM process to be quantified and likely improved, and for the intertwined effect of other processing and APAM to be understood. As well, different applications may have different demands on the APAM material, for example, having lower defect densities or preserving atomic positioning. Most importantly, establishing and then incorporating models of the material behaviour into process simulation tools in technology computer aided design (TCAD) packages can accelerate the development of device process flows. Similarly, the physics underlying different aspects of APAM in device behaviour needs to be studied systematically to understand where device models need to be adapted to account for novel behaviour. Building systematic understanding will require much faster cycles of learning, with the payoff that TCAD-based tool incorporation of this learning enables better device design.

**Advances in Science and Technology to Meet Challenges**

The existing APAM toolset is geared towards the atomic precision required for quantum computing, and not the fast cycles of learning required to develop applications (Figure 2). Fortunately, several recent proofs-of-principle provide a path to a high throughput toolset. Techniques that are capable of surface cleans compatible with APAM have been demonstrated at the wafer scale, including inside a modified chemical vapor deposition (CVD) and, separately, using plasma sputtering. Patterning beyond STM has been demonstrated using electron-beam lithography [8], photolithography [9], and hard masks [3], which enables wafer-scale patterning and overlay alignment. These need to be combined with vacuum tools that do high throughput surface chemistry, such as atomic layer deposition (ALD) or CVD tools in a cluster configuration required to preserve the surface between various processing steps. Involvement by toolmakers can make the community significantly larger by replacing a bespoke toolset with one that feels familiar in a fabrication facility.

Other required advancements are in characterization and modelling. The current patterning tool for APAM, the scanning tunnelling microscope, is also uniquely suited to understand the surface chemistry at the atomic scale but is slow and unreliable for monitoring the process. Alternative techniques for characterizing surfaces, such as auger spectroscopy, x-ray photoelectron spectroscopy, and Fourier-transform infrared spectroscopy, must be explored for faster and large area characterization of the surface. Techniques that can interrogate the function of subsurface APAM structures need to be developed, such as photoluminescence and other optical spectroscopy techniques. After building understanding through characterization, having tools to model and predict behaviour will help drive design and applications. For example, *ab initio* screening with modern high-performance computing could be used to narrow the precursor molecule pool from >100 to 5-10 possibilities for testing. For devices/systems, TCAD tools will require incorporation of new aspects related to non-equilibrium material configurations for process simulation. For TCAD device simulation, work is needed to develop multiscale approaches that can incorporate quantum mechanical details, which may derive from atomic-scale features, into device-scale transport simulations [10].

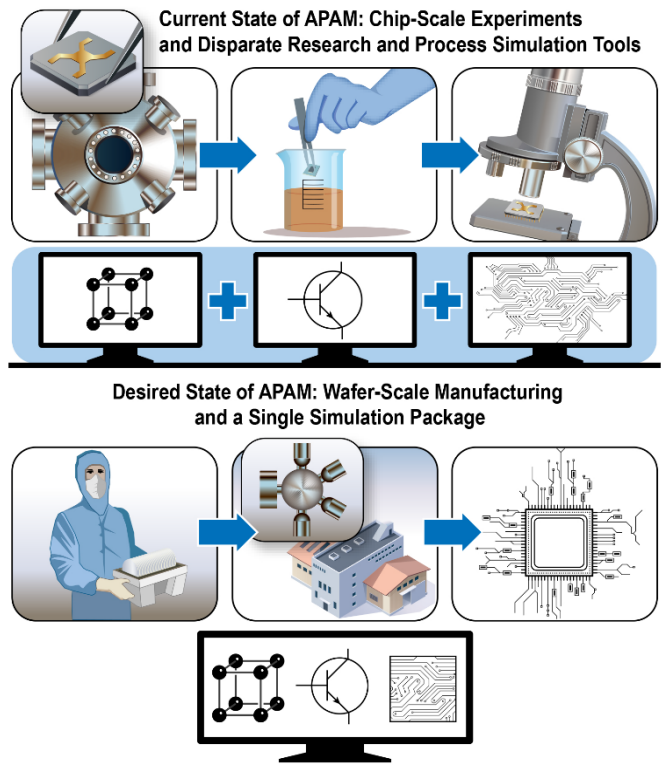

**Figure 2** Future APAM fabrication requires advancing from low-volume experimental and computational tools with high expert intervention to a high-throughput, user-friendly process tool and software package.

**Concluding Remarks**

Current APAM toolsets are built for dopant-based qubit applications, while the material modification afforded by APAM may open a wider application space. The means to incorporate APAM features directly into microelectronic devices and systems have already been established, but transformative applications need to be developed. Aggressive exploration of precursors and characterization of the novel properties of the resultant material can reveal new opportunities. Establishing new integrated toolsets which improve throughput and reliability of the process can enable device development and bring the technique to a wider audience. Deeper understanding of both the relevant material parameters and device physics, and subsequent incorporation of APAM modelling into TCAD tools, can enable device design and performance projections. In total, all the pieces are converging for a disruptive opportunity in materials discovery and associated application development enabled by APAM-like processing.

**Acknowledgements**

We would like to thank Dan Thompson for his artistic vision for the figures in this manuscript. We would also like to thank Chris Arose, DeAnna Campbell, Andrew Leenheer, Denis Mamaluy, Caitlin McCowan, Juan Mendez, David Scrymgeour, Thomas Sheridan, and Thomas Weingartner for their contributions to this work. Support from Sandia's Laboratory Directed Research and Develop Program, the Laboratory for Physical Sciences at the University of Maryland, and the Department of Energy's Advanced Materials and Manufacturing Technologies Office have been critical to our pursuit of work in this area. This work was performed, in part, at the Center for Integrated Nanotechnologies, an Office of Science User Facility operated for the U.S. Department of Energy (DOE) Office of Science. Sandia National Laboratories is a multimission laboratory managed and operated by National Technology & Engineering Solutions of Sandia, LLC, a wholly owned subsidiary of Honeywell International Inc., for the U.S. Department of Energy's National Nuclear Security Administration under contract DE-NA0003525. This paper describes objective technical results and analysis. Any subjective views or